\documentclass[acmsmall,screen,nonacm]{acmart}
\usepackage{multirow}
\usepackage{subcaption}
\usepackage{bbding}
\AtBeginDocument{%
  \providecommand\BibTeX{{%
    \normalfont B\kern-0.5em{\scshape i\kern-0.25em b}\kern-0.8em\TeX}}}

\setcopyright{acmcopyright}
\copyrightyear{2018}
\acmYear{2018}
\acmDOI{10.1145/1122445.1122456}

\begin{document}

\title{Efficient Single Image Super-Resolution Using Dual Path Connections with Multiple Scale Learning}

\author{Bin-Cheng Yang}
\email{yangbincheng@hotmail.com}
\orcid{1234-5678-9012}
\affiliation{%
  \institution{State Key Laboratory for Novel Software Technology, Nanjing University}
  \streetaddress{163 Xianlin Avenue, Qixia District}
  \city{Nanjing}
  \state{Jiangsu}
  \country{China}
  \postcode{210023}
}

\author{Gangshan Wu}
\email{gswu@nju.edu.cn}
\orcid{1234-5678-9012}
\affiliation{%
  \institution{State Key Laboratory for Novel Software Technology, Nanjing University}
  \streetaddress{163 Xianlin Avenue, Qixia District}
  \city{Nanjing}
  \state{Jiangsu}
  \country{China}
  \postcode{210023}
}

\renewcommand{\shortauthors}{Yang and Wu, et al.}

\begin{abstract}
Deep convolutional neural networks have been demonstrated to be effective for single image super-resolution in recent years. On the one hand, residual connections and dense connections have been used widely to ease forward information and backward gradient flows to boost performance.
However, current methods use residual connections and dense connections separately in most network layers in a sub-optimal way. On the other hand, although various networks and methods have been designed to improve computation efficiency, save parameters, or utilize training data of multiple scale factors for each other to boost performance, it either do super-resolution in HR space to have a high computation cost or can not share parameters between models of different scale factors to save parameters and inference time. To tackle these challenges, we propose an efficient single image super-resolution network using dual path connections with multiple scale learning named as EMSRDPN. By introducing dual path connections inspired by Dual Path Networks into EMSRDPN, it uses residual connections and dense connections in an integrated way in most network layers. Dual path connections have the benefits of both reusing common features of residual connections and exploring new features of dense connections to learn a good representation for SISR. To utilize the feature correlation of multiple scale factors, EMSRDPN shares all network units in LR space between different scale factors to learn shared features and only uses a separate reconstruction unit for each scale factor, which can utilize training data of multiple scale factors to help each other to boost performance, meanwhile which can save parameters and support shared inference for multiple scale factors to improve efficiency. Experiments show EMSRDPN achieves better performance and comparable or even better parameter and inference efficiency over state-of-the-art methods. Code will be available at https://github.com/yangbincheng/EMSRDPN.
\end{abstract}



\maketitle

\section{Introduction}
Single image super-resolution (SISR) \cite{DBLP:conf/iccv/FreemanP99} is the task of recovering one high-resolution (HR) image from one low-resolution (LR) image. It has tremendous applications such as medical imaging \cite{DBLP:conf/miccai/ShiCLZBBMDOR13}, satellite imaging \cite{liebel2016single}, surveillance \cite{Zou2010VeryLR} to object recognition \cite{2017EnhanceNet}, which require a lot of image details to help post-precessing. It is an ill-posed problem which has no unique solution because the LR image can be produced from many different HR images. Due to lacking of the prior knowledge of the HR image, it is difficult to recover the HR image from one single LR image.
\begin{figure*}[t]
  \centering
  \begin{minipage}{\linewidth}
  \centering
  \begin{subfigure}{.1344\linewidth}
    \centering\includegraphics[width=\linewidth]{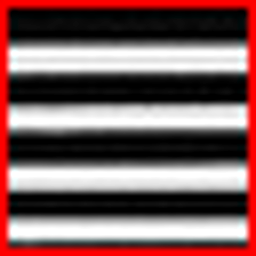}
    \caption*{HR}
  \end{subfigure}
  \begin{subfigure}{.1344\linewidth}
    \centering\includegraphics[width=\linewidth]{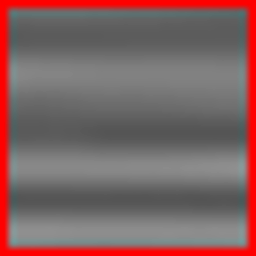}
	\caption*{Bicubic}
  \end{subfigure}
  \begin{subfigure}{.1344\linewidth}
    \centering\includegraphics[width=\linewidth]{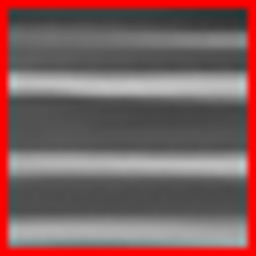}
	\caption*{LapSRN}
  \end{subfigure}
  \begin{subfigure}{.1344\linewidth}
    \centering\includegraphics[width=\linewidth]{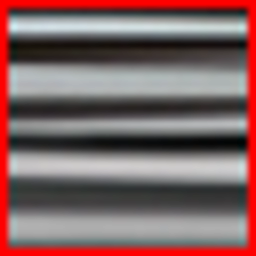}
	\caption*{MDSR}
  \end{subfigure}
  \begin{subfigure}{.1344\linewidth}
    \centering\includegraphics[width=\linewidth]{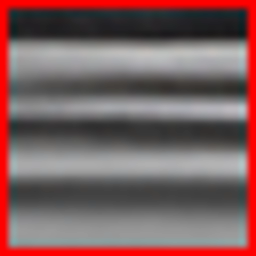}
	\caption*{D-DBPN}
  \end{subfigure}
  \begin{subfigure}{.1344\linewidth}
    \centering\includegraphics[width=\linewidth]{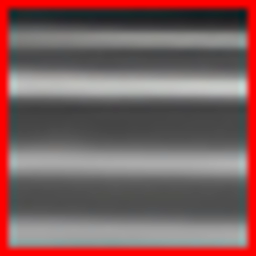}
	\caption*{RDN}
  \end{subfigure}

  \begin{subfigure}{.1344\linewidth}
    \centering\includegraphics[width=\linewidth]{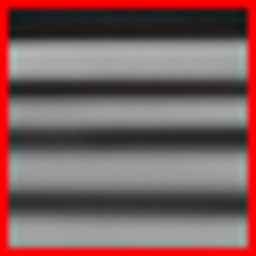}
	\caption*{RCAN}
  \end{subfigure}
  \begin{subfigure}{.1344\linewidth}
    \centering\includegraphics[width=\linewidth]{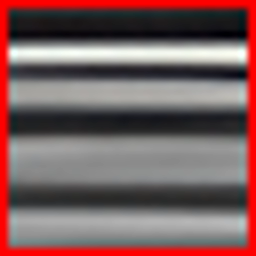}
	\caption*{MSRN}
  \end{subfigure}
  \begin{subfigure}{.1344\linewidth}
    \centering\includegraphics[width=\linewidth]{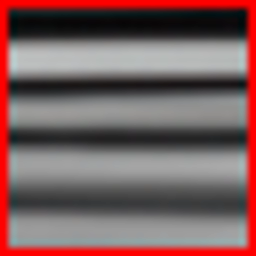}
	\caption*{SAN}
  \end{subfigure}
  \begin{subfigure}{.1344\linewidth}
    \centering\includegraphics[width=\linewidth]{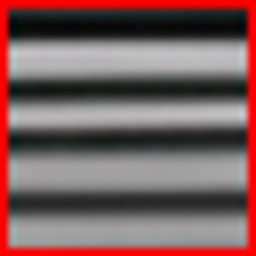}
	\caption*{HAN}
  \end{subfigure}
  \begin{subfigure}{.1344\linewidth}
    \centering\includegraphics[width=\linewidth]{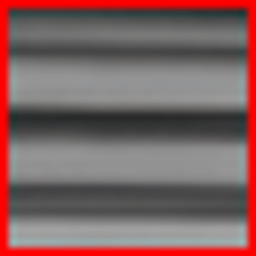}
	\caption*{NLSN}
  \end{subfigure}
  \begin{subfigure}{.1344\linewidth}
    \centering\includegraphics[width=\linewidth]{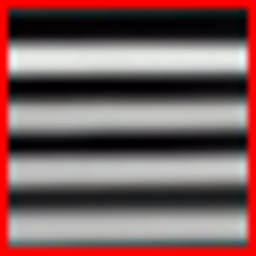}
	\caption*{EMSRDPN}
  \end{subfigure}
  \end{minipage}
  \caption{Zoomed visual results of ``img011'' from Urban100 dataset for $\times{4}$. Our method reconstructs more high frequency details and less artifacts.}
  \label{fig:visual_comparion_zoomed_x4}
\end{figure*}

Many methods have been proposed to tackle this difficult problem, which can be classed into three categories: interpolation based methods \cite{DBLP:conf/icip/AllebachW96,DBLP:journals/tip/LiO01a,Zhang2006AnEI}, model based methods \cite{DBLP:journals/tip/DongZSW11, Zhang2012SingleIS} and example based learning methods \cite{freeman2002example, chang2004super, sun2008image, glasner2009super, yang2010image, dong2016image, zhang2018residual}. Because of the ineffectiveness of interpolation methods for large scale factors and the computation complexity of optimization processes in the prior dependent model based methods, example based learning methods have attracted huge attention to learn a direct mapping from LR image to HR image these years. Especially deep CNN based methods
have been demonstrated to be able to learn the mapping from LR image to HR image effectively.

Since the pioneer work of Dong \MakeLowercase{\textit{et al.}} \cite{dong2016image} first uses a CNN (SRCNN) to solve SISR problem, a lot of deep networks have been proposed to tackle this problem. On the one hand, residual and dense connections are used widely to ease the forward information and backward gradient flows of current deep CNN based methods, which have shown great progress in the reconstruction quality of SISR task. VDSR \cite{kim2016accurate}, DRCN \cite{DBLP:conf/cvpr/KimLL16}, RED \cite{mao2016image}, DRRN \cite{tai2017image}, LapSRN \cite{lai2017deep}, SRGAN/SRResNet \cite{ledig2017photo}, EDSR/MDSR \cite{lim2017enhanced}, RCAN \cite{zhang2018image} and SAN \cite{DBLP:conf/cvpr/DaiCZXZ19} use residual connections global-wise, block-wise, or layer-wise, SRDenseNet \cite{tong2017image}, MemNet \cite{tai2017memnet} and RDN \cite{zhang2018residual} also use dense connections between layers in one block or between different blocks in a network. Although the success of residual and dense connections in SISR networks, they mostly utilize the residual and dense connections separately in most network layers, in this manner which leverage residual and dense connections in a sub-optimal way. On the other hand, how to improve computation efficiency, save parameters, or use training data of multiple scale factors for each other to boost performance is also well studied. Some networks such as VDSR \cite{kim2016accurate}, DRCN \cite{DBLP:conf/cvpr/KimLL16}, DRRN \cite{tai2017image} and MemNet \cite{tai2017memnet} use interpolated LR image as input while sharing all the parameters between different scale factors to leverage training data of multiple scale factors for each other, but they can not do super resolution of multiple scale factors in one pass because of different interpolated inputs for different scale factors and have high computation and memory cost because they do super resolution in interpolated HR space. Other networks such as EDSR \cite{lim2017enhanced}, RDN \cite{zhang2018residual} and RCAN \cite{zhang2018image} use LR image as input directly to train separate models for different scale factors to save computation and memory cost, and use transfer learning to fine-tune large scale factor models from a specific small scale factor model to leverage training data of the specific small scale factor for large scale factors, but they can not share parameters between different scale factors and must do super resolution for different scale factors separately using different models, meanwhile they can not fully leverage training data of multiple scale factors for each other, especially training data of large scale factors for small scale factors.

To overcome these drawbacks, we propose a novel network design named as EMSRDPN by introducing dual path connections and multiple scale training and inference into a very deep convolutional neural network. Through dual path connections, EMSRDPN uses residual and dense connections in an integrated way in most convolutional layers to leverage both residual connections to reuse common features and dense conncetions to explore new features to
learn a good representation for SISR. Meanwhile, by sharing most parameters between different scale factors to use training data of different scale factors for each other to utilize feature correlation during training and amortize most parameters and computation between different scale factors during inference, EMSRDPN has benefits of both achieving better performance as shown in Figure \ref{fig:visual_comparion_zoomed_x4} and improving parameter efficiency and inference time of network.
\begin{figure*}[tp]
\centering\includegraphics[width=\textwidth]{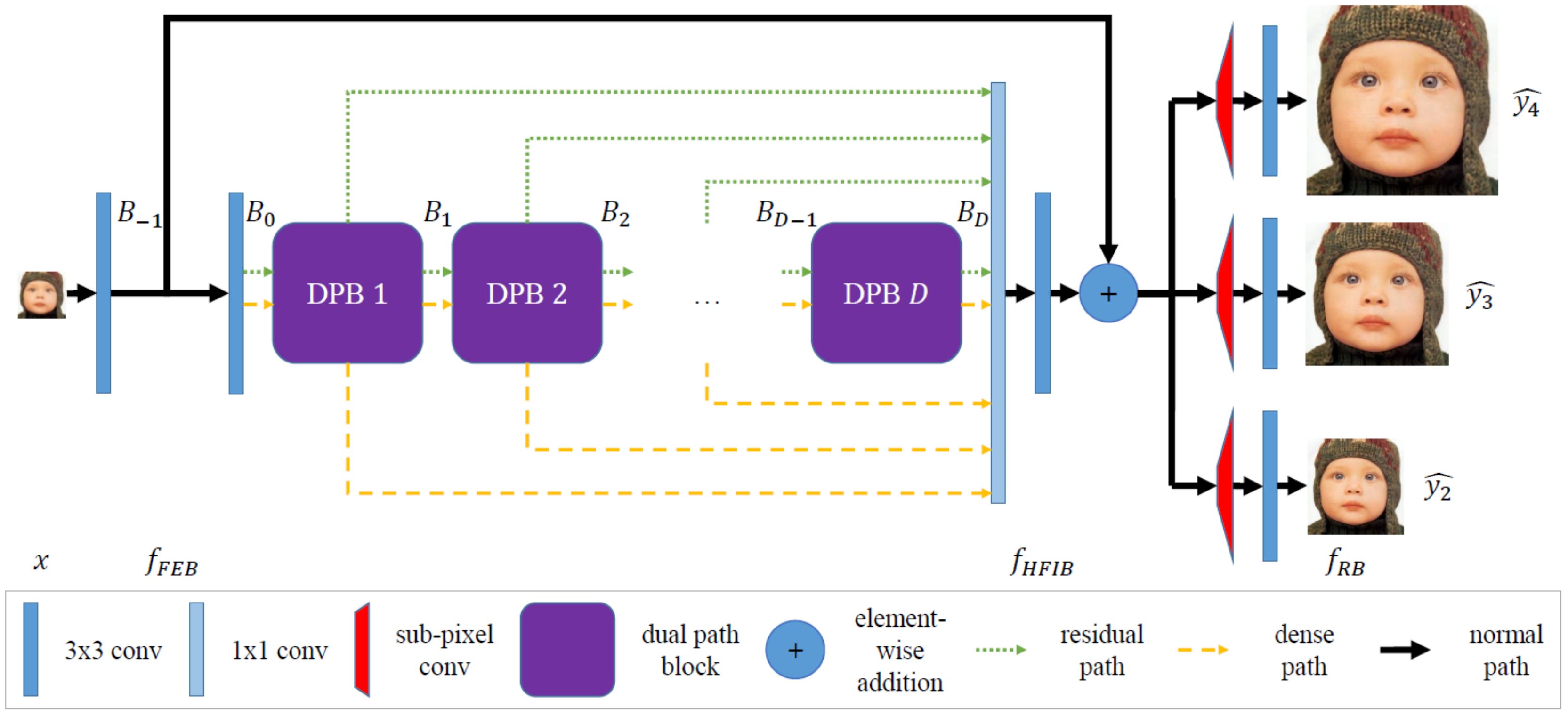}
\caption{Network architecture of EMSRDPN for scale factors $\times{2}$, $\times{3}$, and $\times{4}$.}
\label{fig:architecture_EMSRDPN_EMSRDPN}
\end{figure*}

EMSRDPN includes four parts as shown in Figure \ref{fig:architecture_EMSRDPN_EMSRDPN}, the first part is the feature extraction block (FEB), which extracts low level features from downsampled LR image, the second part is multiple stacked dual path blocks (DPBs) to learn rich hierarchial features from the low level features, each DPB is made of several cascading dual path units (DPUs) and one transition unit, the third part is hierarchical feature integration block (HFIB), which fuses rich hierarchical feature maps output by all the DPBs and FEB both making information forward-propagation direct and alleviating gradient vanishing/exploding problem, the last part is multiple parallel reconstruction blocks (RBs) for different scale factors respectively. FEB, DPBs and HFIB are shared between different scale factors. EMSRDPN is a multi-task super-resolution learning and inference network for multiple scale factors, which can use training data of super-resolution tasks of different scale factors to utilize feature correlation to help each other to improve performance while saving number of parameters and inference time of the model. The key component of network is the dual path unit, which integrates residual connection and dense connections in one layer seamlessly. As pointed by Chen \MakeLowercase{\textit{et al.}} \cite{chen2017dual}, this design can both exploit common features and explore new features to learn rich hierarchical features from low level features, which is beneficial to learn a good representation for SISR. In order to build a deep network, we add a transition unit in every dual path block to fuse residual and dense features to keep computational cost and memory consumption under control.

A preliminary version of this work was presented as a conference paper \cite{10.1145/3343031.3350878}, the model in which is named as SRDPN.
We incorporate additional improvements in significant ways in current work:
\begin{itemize}
\item We propose more efficient dual path units (DPUs) to utilize dual path connections to learn a good representation for SISR, and improve network architecture to do SISR in LR space to save computation and memory cost, these two improvements benefit the efficiency of network and form a base for multiple scale learning.
\item We propose a multiple scale training and inference architecture of network to utilize the feature correlation between multiple scale factors by sharing most of network parameters between different scale factors to make use of training data of multiple scale factors for each other during training and amortize parameters and computation between different scale factors during inference, which has benefits of both achieving better performance and improving parameter efficiency and inference time of network. The improved model in this work is named as EMSRDPN.
\item Extensive experiments on standard benchmark datasets demonstrate the effectiveness of our EMSRDPN model, which achieves better performance and comparable or even better parameter efficiency and inference time compared to state-of-the-art methods.
\end{itemize}
\section{Related Work}
A lot of methods \cite{DBLP:conf/icip/AllebachW96, DBLP:journals/tip/LiO01a, Zhang2006AnEI, freeman2002example, yang2010image, sun2008image, tai2010super, chang2004super, DBLP:journals/tip/DongZSW11, Zhang2012SingleIS, timofte2014a+, zeyde2010single, yang2013fast2, glasner2009super, wang2015deep, DBLP:conf/cvpr/ZhangZ018} have been proposed to solve classic SISR problem for decades. We focus on recent CNN based methods for SISR in this section. We first introduce development of CNN based methods for SISR. Second we discuss current use of residual and dense connections in CNN based methods, their limitation and our motivation. Thirdly, we discuss multiple scale training and inference strategies of current CNN based methods, their limitations and our motivation.
\subsection{CNN Based Methods for SISR}

The pioneer work of Dong \MakeLowercase{\textit{et al.}} \cite{dong2016image} first uses a CNN (SRCNN) to solve SISR problem which processes interpolated LR image and learns feature extraction, feature mapping and HR image reconstruction jointly. Furthermore, Dong \MakeLowercase{\textit{et al.}} \cite{dong2016accelerating} improve SRCNN by designing a deeper network architecture (FSRCNN) which directly processes LR image and uses a deconvolution layer to do upscaling. To improve the computation efficiency further, Shi \MakeLowercase{\textit{et al.}} \cite{shi2016real} introduce an efficient sub-pixel convolution layer to do HR image reconstruction (ESPCN). From a human perception view, Johnson \MakeLowercase{\textit{et al.}} \cite{johnson2016perceptual} propose a perceptual loss function based on IMAGENET \cite{Deng2009ImageNetAL} pre-trained features. To use one model to handle SISR for different scale factors, Kim \MakeLowercase{\textit{et al.}} \cite{kim2016accurate} propose a \(20\)-layer convolution network (VDSR) to leverage large context regions, and use residual learning and adjustable gradient clipping to ease training process. Moreover, to improve the model parameter efficiency, they also propose a deeply-recursive convolutional network (DRCN) \cite{DBLP:conf/cvpr/KimLL16} using recursive supervision and skip connections. Afterwards, Tai \MakeLowercase{\textit{et al.}} \cite{tai2017image} adopt both global and local residual leaning into recursive network (DRRN) to mitigating training difficulty of network while reducing the model complexity. Also to implement a single model which can process multiple scale factors, Lai \MakeLowercase{\textit{et al.}} \cite{lai2017deep} propose a Laplacian Pyramid Super Resolution Network (LapSRN) to progressively reconstruct sub-bands of the HR image, it can produce HR images of different scale factors in one pass. To leverage relations between SISR tasks of different scale factors, Lim \MakeLowercase{\textit{et al.}} \cite{lim2017enhanced} first propose enhanced deep residual networks (EDSR) for SISR which remove unnecessary blocks in residual networks and adopt residual scaling, then they use transfer learning from small scale factor models to large scale factor models. Moreover, to exploit inter-task correlation of different scale factors, they design a multi-scale architecture (MDSR) based on EDSR which uses the common network trunk and different network heads and tails for SISR tasks of different scale factors. From different points of view, Hui \MakeLowercase{\textit{et al.}} \cite{hui2018fast} design an information distillation network (IDN) to reduce computation cost and memory consumption, Haris \MakeLowercase{\textit{et al.}} \cite{haris2018deep} propose a deep convolutional neural network (DBPN) to implement iterative back projection algorithm \cite{Irani1991ImprovingRB}, Li \MakeLowercase{\textit{et al.}} \cite{li2018multi} design a multi-scale residual network (MSRN) to detect features at different scales and combine them to boost performance, and Qiu \MakeLowercase{\textit{et al.}} propose a novel embedded block residual network (EBRN) \cite{DBLP:conf/iccv/QiuWT019} which uses different modules to restore information of different frequencies.

Some networks focus on using residual connections and dense connections to boost performance. Mao \MakeLowercase{\textit{et al.}} \cite{mao2016image} use a symmetric convolution-deconvolution architecture (RED) and skip connections to recursively learn the residual of HR image and LR image. Then Ledig \MakeLowercase{\textit{et al.}} \cite{ledig2017photo} first use adversarial learning to generate HR image (SRGAN) based on a residual network based generator network (SRResNet), which uses residual blocks and global residual learning to reconstruct HR image. On the other hand, Tong \MakeLowercase{\textit{et al.}} \cite{tong2017image} introduce dense connections to a very deep network for SISR (SRDenseNet), both enabling an effective way to combine features and alleviating the gradient vanishing/exploding problem. Moreover, Tai \MakeLowercase{\textit{et al.}} \cite{tai2017memnet} propose a persist memory network design (MemNet) to leverage long-term and short-term memories, which uses dense connections among big blocks in a global way. Afterwards, Zhang \MakeLowercase{\textit{et al.}} \cite{zhang2018residual} propose residual dense network (RDN) to better exploit multi-level features in the network, which uses dense connections among the layers in local blocks and residual connections in blocks and globally. And Anwar \MakeLowercase{\textit{et al.}} \cite{Anwar2022DenselyRL} propose densely connected residual blocks and an attention network (DRLN) to utilize residual connection, dense connection and attention for SISR.

Attention mechanisms and non-local operations have also been introduced into deep CNN based methods to improve perfomance further. Zhang \MakeLowercase{\textit{et al.}} further introduce attention mechanism in convolutional network (RCAN) \cite{zhang2018image} to solve SISR, which uses skip connections in multiple levels. After that, Dai \MakeLowercase{\textit{et al.}} propose a second-order attention network (SAN) \cite{DBLP:conf/cvpr/DaiCZXZ19} with residual connections for more powerful feature expression and feature correlation learning. Then Zhang \MakeLowercase{\textit{et al.}} propose kernel attention network \cite{zhang2020kernel} for SISR by dynamically selecting appropriate kernel size to adjust receipt field size, Niu \MakeLowercase{\textit{et al.}} \cite{niu2020single} propose HAN to model the holistic interdependencies among layers, channels and positions, Li \MakeLowercase{\textit{et al.}} propose a lightweight dense connection distillation network \cite{li2021lightweight} by incorporating contrast-aware channel attention. Since Liu \MakeLowercase{\textit{et al.}} \cite{DBLP:conf/nips/LiuWFLH18} first introduce non-local operations into a recurrent neural network (NLRN) to capture feature correlation and improve parameter efficiency for image restoration, Zhang \MakeLowercase{\textit{et al.}} \cite{zhang2019residual} propose RNAN to use local and non-local attention blocks to capture long-range dependency and attend to challenging parts and Mei \MakeLowercase{\textit{et al.}} \cite{mei2021image} combine non-local operation and sparse representation into deep networks (NLSN) to boost performance and efficiency.

\subsection{Residual and Dense Connections Used in SISR Networks}
Skip connections, especially residual connections and dense connections, are adopted widely to make use of hierarchical features and ease training of very deep networks. On the one hand, VDSR \cite{kim2016accurate}, DRCN \cite{DBLP:conf/cvpr/KimLL16}, DRRN \cite{tai2017image}, MemNet \cite{tai2017memnet}, EDSR\cite{lim2017enhanced}, LapSRN \cite{lai2017deep}, DBPN \cite{haris2018deep}, RED \cite{mao2016image}, SRResNet \cite{ledig2017photo}, RDN \cite{zhang2018residual}, RCAN \cite{zhang2018image} and SAN \cite{DBLP:conf/cvpr/DaiCZXZ19} all use residual connections to ease forward information flow and backward gradient flow of networks to boost performance. On the other hand, MemNet \cite{tai2017memnet}, DBPN \cite{haris2018deep}, SRDenseNet \cite{tong2017image} and RDN \cite{zhang2018residual} also use dense connections to leverage multi-level features and alleviate the gradient vanishing/exploding problem to improve performance.

Although residual connections and dense connections have been used to boost the SR performance of deep CNN based methods effectively, the current methods only use residual connections and dense connections in a separate way in most network layers. To learn a good representation for single image super-resolution by using residual connections and dense connections in a better way, we introduce dual path connections inspired by \cite{chen2017dual} into a very deep neural network, which has the benefits of both residual connections to reuse common features and dense connections to explore new features to learn good features.

\subsection{Multiple scale Training and Inference for SISR Networks}
Since VDSR \cite{kim2016accurate} first uses the interpolated LR image as input to the network to leverage training data of multiple scale factors for each other and share model parameters between different scale factors, DRCN \cite{DBLP:conf/cvpr/KimLL16}, DRRN \cite{tai2017image} and MemNet \cite{tai2017memnet} all follow this strategy. Although this kind of strategy has the benefits of both exploiting correlation between SISR task of different scale factors to boost performance and sharing one model for different scale factors to improve parameter efficiency, it has two limitations. First, this strategy does super resolution in the HR space which has high computation and memory cost. Second, it can not do multi-scale super resolution because the interpolated inputs to the network for different scale factors will be different, leading to multiple passes of inference for different scale factors. On the other hand, EDSR \cite{lim2017enhanced}, RDN \cite{zhang2018residual} and RCAN \cite{zhang2018image} use transfer learning to fine-tune model parameters of large scale factors from model parameters of a specific small scale factor. Although this kind of strategy somehow utilizes the knowledge of small scale factor task to boost the performance of large scale factor tasks, it has some limitations. First, the training data of different scale factors can not be utilized for each other, especially training data of large scale factors for small scale factors. Second, the parameters thereafter models can not be shared between different scale factors impacting the parameter efficiency because the parameters are different for different scale factors after fine-tuning. At last, this strategy can not do multi-scale super resolution either because different scale factors have different parameters and models although the inputs to different models can be the same. Besides these two main kind of strategies, a variety of EDSR \cite{lim2017enhanced} model named MDSR \cite{lim2017enhanced} utilizes shared parameters of network trunk between different scale factors and uses training data of different scale factors to help each other, but it has limitation. Although the parameters of network trunk are shared, MDSR has different pre-processing and upsampling modules for different scale factors, leading to which can not do multiple scale super resolution because there will be different input feature maps to shared network trunk after pre-processing by different pre-processing modules during inference. LapSRN \cite{lai2017deep} and MS-LapSRN \cite{lai2018fast} also use training data of different scale factors to help each other and share parameters between different scale factors, their strategy is different from ours which need more inference stages therefore more parameter, computation and memory budgets for large scale factors.

To overcome these drawbacks and utilize feature correlation between multiple scale factors, we propose a multiple scale training and inference architecture of network which share most of network parameters and use only different parameters in network tails for different scale factors, which have benefits of shared parameters and shared inference in most of network to boost efficiency and using training data of different scale factors to train shared features to boost performance.

\section{Method}
In this section, we describe network architecture and components of EMSRDPN in detail.
\subsection{Network Architecture}
The proposed EMSRDPN, as shown in Figure \ref{fig:architecture_EMSRDPN_EMSRDPN}, consists of four parts: a feature extraction block (FEB), multiple stacked dual path blocks (DPBs), a hierarchical feature integration block (HFIB), and finally multiple parallel reconstruction blocks (RBs) for different scale factors. Let's denote \(x\) as the input and \(\hat{y_{s}}\) as the output of the network for scale factor \(s\), which is arbitrary one of the different scale factors of network without loss of generality. First, we use two convolution layers in FEB to extract low level features from network input \(x\),
\begin{equation}B_{0}=f_{FEB}(x),\end{equation}
where \(f_{FEB}\) denotes function of FEB, \(B_{0}\) denotes the extracted features to be sent to the first DPB and \(x\) denotes downsampled LR image which will be upscaled. Second, we use multiple stacked DPBs to learn rich hierarchical features from low level features. Suppose \(D\) blocks are stacked, we have
\begin{equation}B_{d} = f_{DPB,d}(f_{DPB,d-1}(...(f_{DPB,1}(B_{0}))...)), d = 1,2,...,D,\end{equation} where \(f_{DPB,d}\) denotes the function of \(d\)-th DPB, \(B_{d}\) denote the output feature maps of the \(d\)-th DPB. Afterwards, we integrate rich hierarchical feature maps output by all the DPBs and FEB using HFIB,
\begin{equation}{H} = f_{HFIB}(B_{-1}, [B_{1}, B_{2}, ..., B_{D}]),\end{equation}
where \(B_{-1}\) denotes the output of first convolution layer of FEB, \(f_{HFIB}\) denotes the function of HFIB and \(H\) denotes the output of HFIB.
Finally we use integrated features output by HFIB as input to RB specific to scale factor \(s\) to reconstruct the HR image \(y_{s}\),
\begin{equation}\hat{y_{s}} = g_{s}(x) = f_{RB_{s}}({H}),\end{equation}
where \(f_{RB_{s}}\) denotes the function of the RB for scale factor \(s\) and \(g_{s}\) denotes the function of EMSRDPN for scale factor \(s\) respectively. (In this paper, either \([B_{1}, B{2}, ..., B_{D}]\) denotes the concatenation of feature maps along feature dimension when used on the right-side of formulae or as the arguments of functions, or \([B_{1}, B{2}, ..., B_{D}]\) denotes the split of feature maps along feature dimension when used on the left-side of formulae.)

Given a set of training image pairs \(\{x^{(k)}_{s}, y^{(k)}\}^{M}_{k=1}, s \in S\) for a set of scale factors \(S\), the network is used to minimize the following Mean Absolute Error (MAE) loss for every scale factor \(s \in S\) alternately:
\begin{equation}L(\Theta_{s})=\frac{1}{M}\sum^{M}_{k=1}||y^{(k)}-\hat{y}^{(k)}_{s}||_{1},\end{equation}
where \(\Theta_{s}\) is the parameters of EMSRDPN for scale factor \(s\), which includes shared parameters in FEB, DPBs, HFIB for all the different scale factors and private parameters in RB corresponding to scale factor \(s\). Note that we have only one HR image and different LR images for different scale factors during training using standard training datasets but we can infer multiple SR images from one LR image during testing.

In the following subsections, we will describe feature extraction block (FEB), dual path block (DPB), hierarchical feature integration block (HFIB), and reconstruction block (RB) in detail.
\subsection{Feature Extraction Block}
The FEB uses two \({3\times3}\) convolution layers to extract low level features from the LR image \(x\). To be consistent with DPBs, we also take it's output as two parts, the residual part and the dense part,
\begin{equation}
B_{-1} = W^{1}_{F}(x),
\end{equation}
\begin{equation}
f_{FEB}(x) = B_{0} = W^{2}_{F}(B_{-1}),
\end{equation}
\begin{equation}
[R^{0}_{1}, D^{0}_{1}] = B_{0},
\end{equation}
where \(W^{1}_{F}\) and \(W^{2}_{F}\) denote the wights of the first and second convolution layer in FEB (we omit bias for simplicity in this section and following sections), \(B_{-1}\) denotes the output of first convolution layer in FEB, \(B_{0}\) denotes the output of second convolution layer in FEB, \(R^{0}_{1}\) and \(D^{0}_{1}\) denote the residual part and the dense part of output \(B_{0}\) after splitting along feature dimension which will be the input to the first DPB.

\subsection{Dual Path Block}
\begin{figure}[t]
\begin{center}
   \includegraphics[width=.56\linewidth]{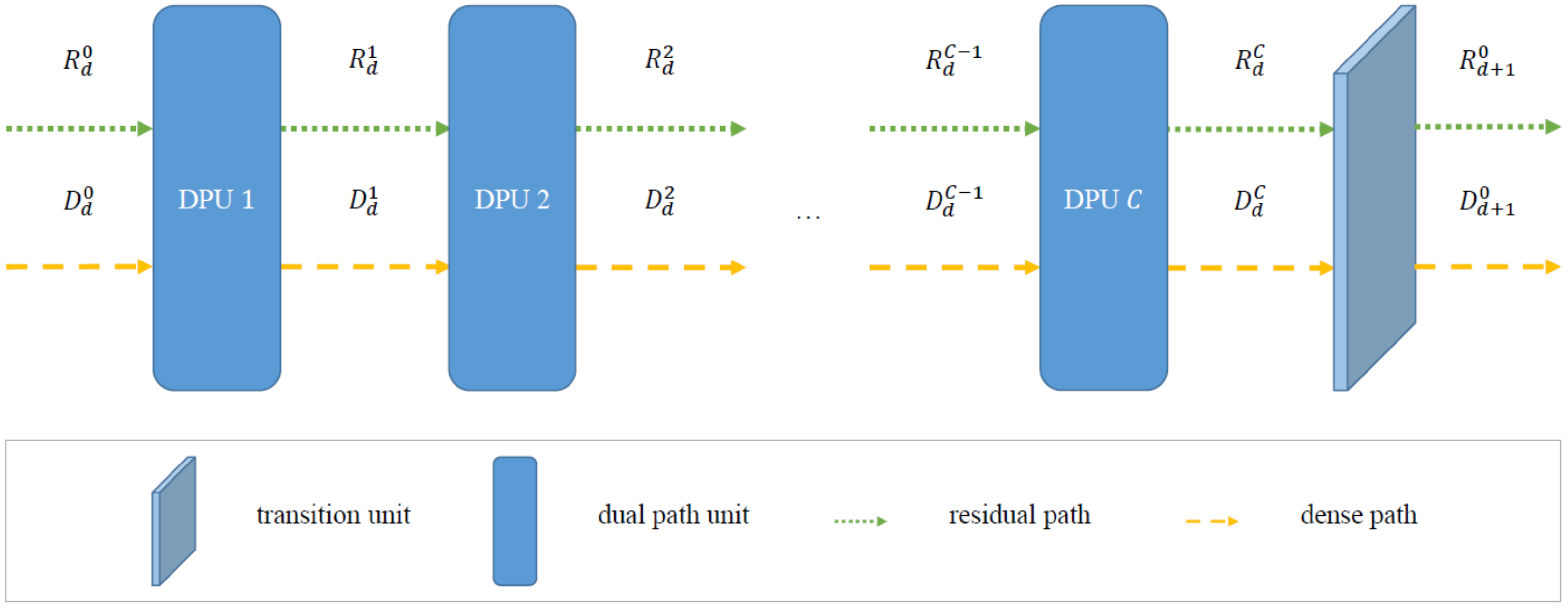}
\end{center}
   \caption{The structure of \(d\)th DPB.}
\label{fig:dpb}
\end{figure}

As shown in Figure \ref{fig:dpb}, each DPB is composed of several cascading dual path units (DPUs) and one transition unit (TU). Let's assume our network has \(D\) DPBs and each DPB has \(C\) DPUs. The input feature maps of \(d\)th DPB is the output feature maps of \((d-1)\)th DPB or the output feature maps of FEB when \(d=1\). Let's denote \(R^{c}_{d}\) and \(D^{c}_{d}\) as the residual part and the dense part of output feature maps of \(c\)th DPU in \(d\)th DPB or the residual part and the dense part of output feature maps of TU in \((d-1)\)th DPB when \(c=0\), \(d\neq1\) or the residual part and the dense part of output feature maps of FEB when \(c=0\), \(d=1\).

\subsubsection{Dual Path Unit}
\begin{figure}[t]
\begin{center}
   \includegraphics[width=.56\linewidth]{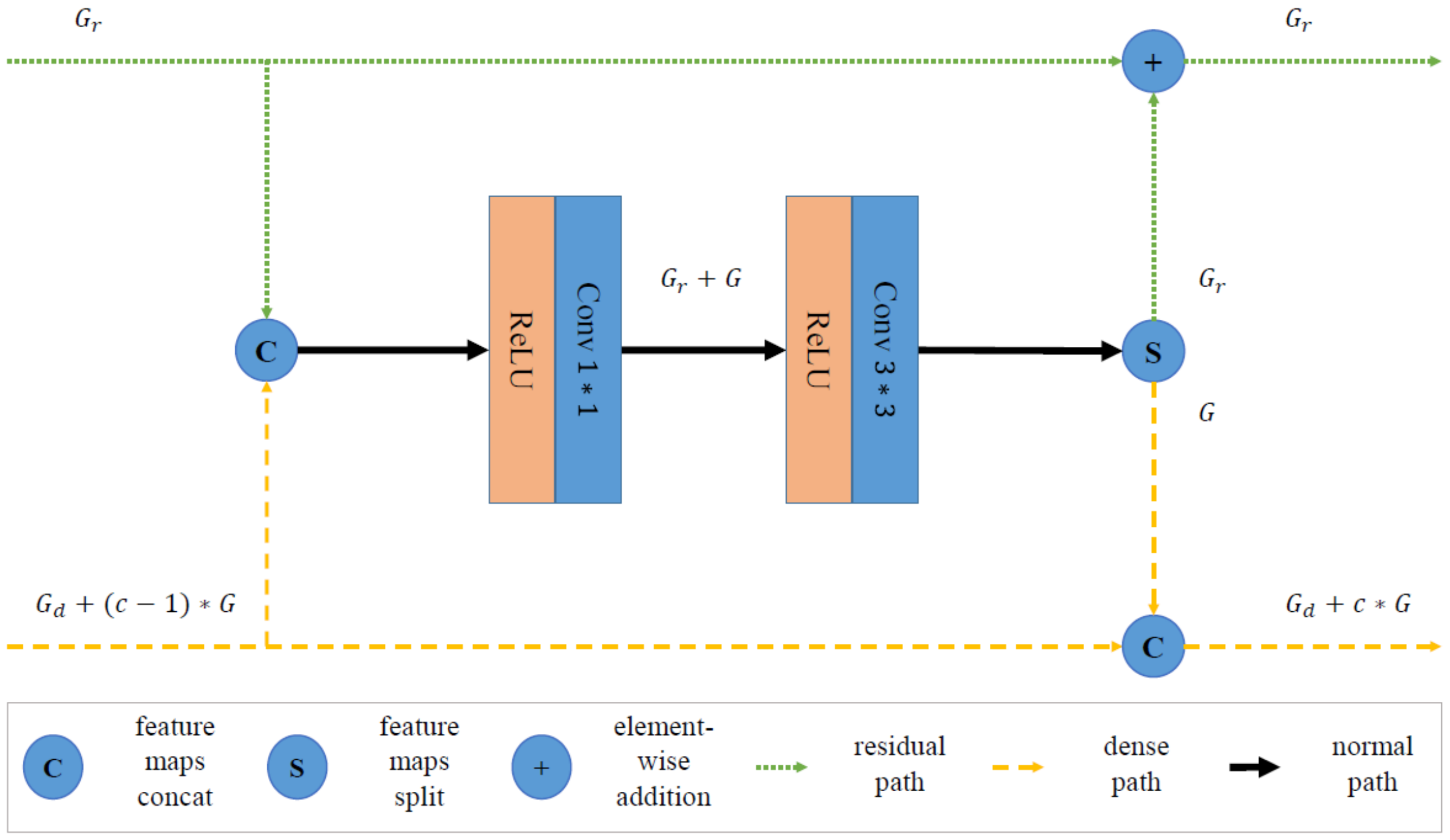}
\end{center}
   \caption{The structure of \(c\)th DPU.}
\label{fig:dpl}
\end{figure}

DPU is shown in Figure \ref{fig:dpl}. Each DPU takes it's input feature maps as two parts, the residual part and the dense part. The input of first DPU in \(d\)th DPB is the output of transition unit in \((d-1)\)th DPB or the output of FEB when \(d=1\), which is denoted as \(R^{0}_{d}\) and \(D^{0}_{d}\). The input of \(c\)th DPU except the first DPU in \(d\)th DPB is the output of \((c-1)\)th DPU in \(d\)th DPB.

Each DPU first concatenates the residual part and the dense part of its input feature maps, then does non-linear transform to the concatenated input feature maps because we adopt the ``\textit{pre-activation}'' design of residual network \cite{he2016identity}, at last it uses a bottleneck design which includes a \(1\times1\) convolution layer followed by a \(3\times3\) convolution layer to learn a residual function and new dense features while keeping the computation and memory budget low,
\begin{equation}
C^{c}_{d} = W^{c}_{d}(\sigma(T^{c}_{d}(\sigma([R^{c-1}_{d}, D^{c-1}_{d}])))),
\end{equation}
where \(T^{c}_{d}\) and \(W^{c}_{d}\) denote the weights of the \(1\times1\) convolution layer and \(3\times3\) convolution layer of \(c\)th DPU in \(d\)th DPB, \(\sigma\) denotes nonlinear function, \(C^{c}_{d}\) denotes the output of the \(3\times3\) convolution layer in this DPU. Next it splits the output into two parts, a residual function to be added to the residual part of input and new dense features to be concatenated to the dense part of input,
\begin{equation}
[F^{c}_{d}, G^{c}_{d}] = C^{c}_{d},
\end{equation}
where \(F^{c}_{d}\) denotes the residual function to be added to the residual part of input feature maps and \(G^{c}_{d}\) denotes newly produced dense features to be concatenated to the dense part of input feature maps. Finally, it add the residual function to the residual part of input feature maps to form ``\textit{residual path}'',
\begin{equation}
R^{c}_{d} = R^{c-1}_{d} + F^{c}_{d},
\end{equation}
and concatenate the newly learned dense features to the dense part of the input feature maps to form ``\textit{dense path}'',
\begin{equation}
D^{c}_{d} = [D^{c-1}_{d}, G^{c}_{d}].
\end{equation}
These two paths of feature maps are used as input to the next DPU in current DPB or as input to the TU in current DPB when this DPU is the last DPU in current DPB. Batch normalization layer is removed according to \cite{lim2017enhanced} and and non-linearity transform is changed to ReLU to improve computation and memory efficiency in this work compared to SRDPN.
\begin{figure}[t]
\begin{center}
   \includegraphics[width=.56\linewidth]{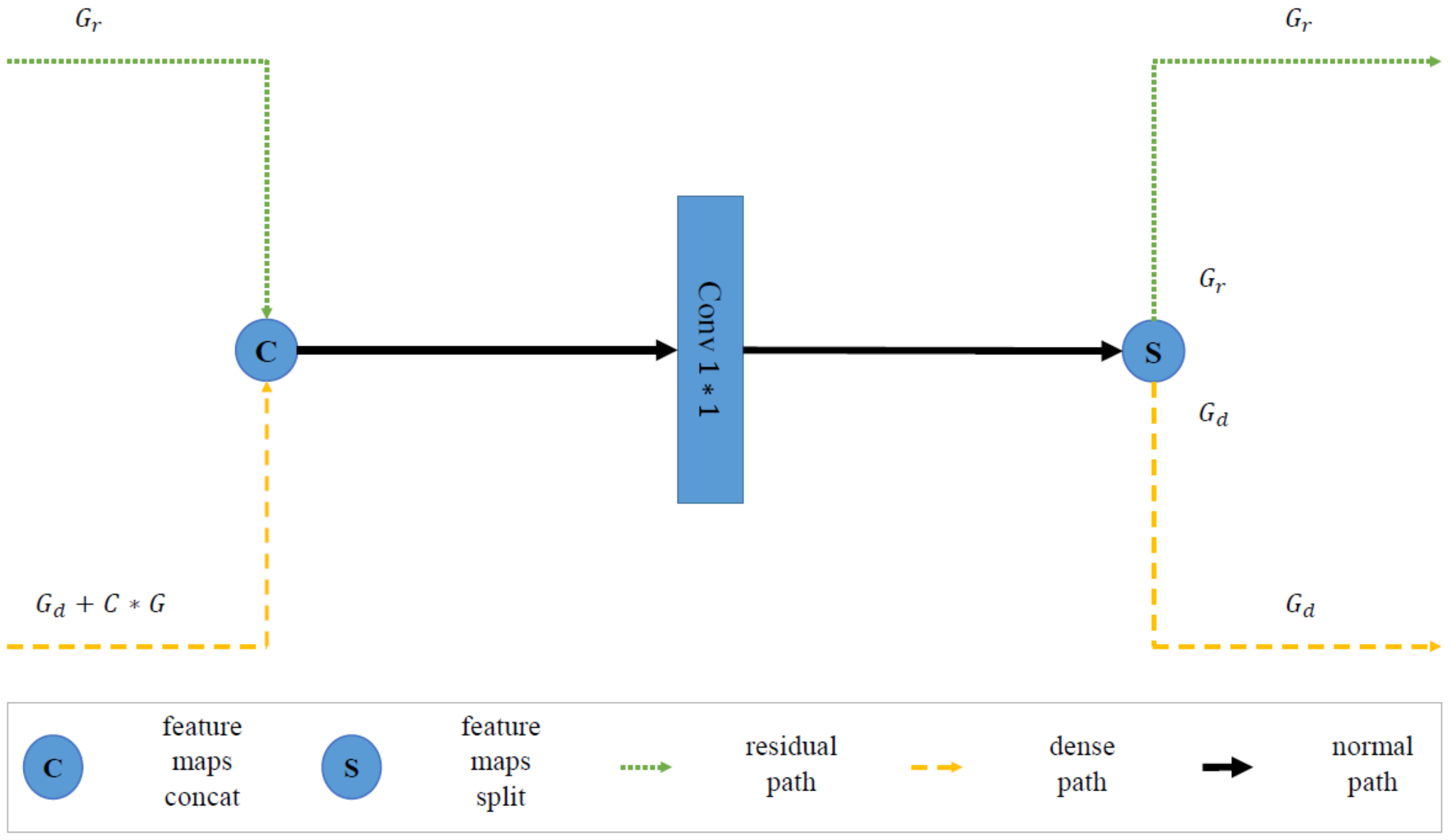}
\end{center}
   \caption{The structure of TU.}
\label{fig:dpl_TU}
\end{figure}
\subsubsection{Transition Unit}
Because the dense path of each DPU in every DPB will concatenate dense part of it's output to dense part of it's input feature maps, which is all the dense parts of outputs of previous DPUs and dense part of input of the first DPU in current DPB. Assuming the input feature maps to first DPU has \(G_{r}\) residual path features and \(G_{d}\) dense path features, and the grow rate of dense path of each DPU is \(G\), after the \(C\)th DPU of \(d\)th DPB, we have \(G_{r}\) residual path features and \((G_{d}+C\times{G})\) dense path features which are the input of \((d+1)\) DPB. In order to build deeper EMSRDPN and make computation complexity and memory consumption under control, we add a transition unit in the tail of each DPB to transform all the \((G_{r}+G_{d}+C\times{G})\) output features of current DPB to \((G_{r}+G_{d})\) features to be the input to the first DPU of next DPB. Each transition unit also takes its input feature maps as two parts, the residual part and the dense part, which are output by the last DPU in current DPB. The structure of TU is shown in Figure \ref{fig:dpl_TU}, the function of TU in \(d\)th DPB is formulated as follows:
\begin{equation}
f_{DPB,d}(B_{d-1}) = B_{d} = W^{T}_{d}([R^{C}_{d-1},D^{C}_{d-1}]),
\end{equation}
\begin{equation}
[R^{0}_{d+1}, D^{0}_{d+1}] = B_{d},
\end{equation}
where \(W^{T}_{d}\) denotes the weights of \(1\times1\) convolutional layer in TU, \(B_{d}\) denotes the output of convolution layer in TU, \(R^{0}_{d+1}\) and \(D^{0}_{d+1}\) denote the residual part and dense part of output feature maps of TU after splitting along feature dimension, which are the input of first DPU in next DPB. The output of \(d\)th DPB is the output of the TU in \(d\)th DPB.

\subsection{Hierarchical Feature Integration Block}
In hierarchical feature integration block, we first concatenate outputs of all the DPBs in order to use rich hierarchical features learned by multiple stacked DPBs,
\begin{equation}F = [B_{1},B_{2},...,B_{D}],\end{equation}
then use a \(1\times1\) convolution layer to transform the feature space to low dimensionality to reduce computation cost and memory consumption,
and finally use a \(3\times3\) convolution layer to learn a global residual function to be added to the output of the first convolution layer in FEB of network (denoted as \(B_{-1}\)) to improve forward information and backward gradient flows further following the strategy of Zhang \MakeLowercase{\textit{et al.}} \cite{zhang2018residual},
\begin{equation}H = f_{HFIB}(B_{-1},F) = W^{2}_{H}(W^{1}_{H}(F)) + B_{-1},\end{equation}
where \(W^{1}_{H}\) and \(W^{2}_{H}\) denote the weights of \(1\times1\) convolution layer and \(3\times3\) convolution layer in HFIB respectively, \(H\) denotes the output of HFIB.

\subsection{Reconstruction Block}
In order to use training data of multiple scale factors for each other to utilize featrue correlation to boost performance and improve parameter efficiency and inference time of the proposed model further, we use multiple parallel RBs for different scale factors respectively and shared FEB, DPBs and HFIB for different scale factors in EMSRDPN, which can produce HR images of different scale factors in one pass. The number of and corresponding scale factors of RBs can be configured flexibly according to availability of training data and requirements of applications. In the reconstruction block (RB) for a specific scale factor \(s\), we use the output of hierarchical feature integration block (HFIB) shared by different scale factors as the input to an efficient sub-pixel convolution layer proposed by \cite{shi2016real} followed by a \(3\times3\) convolution layer to reconstruct the HR image for scale factor \(s\),
\begin{equation}\hat{y_{s}} = f_{RB_{s}}(H) = W^{2}_{R_{s}}(W^{1}_{R_{s}}(H)),\end{equation}
where \(W^{1}_{R_{s}}\) and \(W^{2}_{R_{s}}\) denote the weights of efficient sub-pixel convolution layer and \(3\times3\) convolution layer in RB for scale factor \(s\) respectively.

\section{Experiments}
In this section, we first discuss training and testing data sets and metrics for performance evaluation. Next we describe implementation details of the proposed EMSRDPN. Then we study effects to reconstruction performance of different number of scale factors for training. Furthermore, we compare the performance of EMSRDPN with the state-of-the-art CNN based SISR methods. Finally, we investigate the model complexity and inference time of EMSRDPN.
\subsection{Datasets and Evaluation Metrics}
Following the setting in \cite{haris2018deep, zhang2020residual}, we use DIV2K \cite{Agustsson2017NTIRE2C} and Flickr2K \cite{Lim_2017_CVPR_Workshops} as the training sets, five standard benchmark data sets, Set5 \cite{bevilacqua2012low}, Set14 \cite{zeyde2010single}, BSD100 \cite{DBLP:conf/iccv/MartinFTM01}, Urban100 \cite{huang2015single} and Manga109 \cite{DBLP:conf/icpr/FujimotoOYMYA16}, are used for performance evaluation. Becuase some methods only use DIV2K \cite{Agustsson2017NTIRE2C} as the training set, we also train a model denoted as EMSRDPN-D using only DIV2K \cite{Agustsson2017NTIRE2C} for reference and comparison.
DIV2K and Flickr2K datasets provide corresponding HR images and LR images which are downsampled based on bicubic interpolation. We do two kinds of data augmentation during training: 1) Rotate the images of 90, 180, 270 degrees; 2) Flip the images vertically. We use peak signal-to-noise ratio (PSNR) and structural similarity index (SSIM) as evaluation metrics. We train the network using all three RGB channels and test the network from Y channel of YCbCr color space of images following the strategy of most state-of-the-art methods.
\subsection{Implementation Details}
  We construct a network with (\(D=16\)) DPBs and each DPB has (\(C=4\)) DPUs. The width of residual path of network is set to (\(G_{r}=64\)), the width of basic dense path of network is set to (\(G_{d}=64\)) which is the width of dense part of TU and FEB, and the grow rate for the dense path of network is set to (\(G=64\)) which is the width of dense part of DPU. The width of output feature maps of convolutional layers in FEB and RBs is set to (\(G_{r}+G_{d}\)), except that the output of the last \(3\times3\) convolutional layers in RBs is multiple reconstructed HR color images corresponding to different scale factors. All layers use \(3\times3\) convolutional kernels except transition layers in DPUs, TUs and HFIB, which use \(1\times1\) convolutional kernels to fuse features and reduce computation and memory cost. Compared to SRDPN, rectifier linear unit \cite{DBLP:conf/iwaenc/XuCL16} is used as non-linear function of our network and batch normalization \cite{pmlr-v37-ioffe15} is removed before non-linearity. We implement our EMSRDPN with PyTorch framework. Mini-batch size is set to \(16\) due to memory limit. We use \(48\times48\) LR image patches and corresponding HR image patches for \(\times2\), \(\times3\), \(\times4\) and \(\times8\) scale factors to train our models. At each iteration of training EMSRDPN, we randomly sample a scale factor uniformly to sample mini-batch training pairs of this scale factor to update shared parameters in FEB, DPBs, HFIB and private parameters in RB for the sampled scale factor. For a sampled specific scale factor, we first uniformly sample an image pair from all pairs of corresponding HR and LR images, then uniformly sample a training patch pair from the sampled image pair. Adam optimizer \cite{DBLP:journals/corr/KingmaB14} is used to learn the network weights, learning rate is initialized to \(1\times10^{-4}\). We train our network for about \(1\times10^{6}\) iterations on average for each scale factor in set of all training scale factors, the learning rate is step-decayed by \(2\) after about number of training scale factors times \(2\times10^{5}\) of iterations for multiple scale training. We implement our method using PyTorch framework, train our models using NVIDIA Tesla V100 GPUs and test our models using NVIDIA TITAN Xp GPUs.
\begin{table*}[tb]
{
\fontsize{5}{6}\selectfont
\captionsetup{justification=centering}
\caption{Average PSNR/SSIM Metrics for Scale Factors \(\times2\), \(\times3\), \(\times4\), \(\times8\) of EMSRDPN Using Different Number of Training Scale Factors. Best is \textbf{Highlighted}, and Second Best is \underline{Underlined.}
}
\begin{center}
{
\begin{tabular}{|c|c|c|c|c|c|c|c|c|c|c|c|}
\hline
\multirow{2}*{Method}&\multirow{2}*{Scale}&\multicolumn{2}{|c|}{Set5}&\multicolumn{2}{|c|}{Set14}&\multicolumn{2}{|c|}{B100}&\multicolumn{2}{|c|}{Urban100}&\multicolumn{2}{|c|}{Manga109}\\
\cline{3-12}
&&PSNR&SSIM&PSNR&SSIM&PSNR&SSIM&PSNR&SSIM&PSNR&SSIM\\
\hline
\hline
EMSRDPN-2 &$\times2$&\underline{38.24}&0.9658&33.96&0.9291&32.35&0.9115&32.95&0.9407&39.38&0.9801\\
EMSRDPN-23 &$\times2$&38.23&0.9658&34.17&\underline{0.9311}&32.39&0.9118&33.20&0.9424&\underline{39.48}&\textbf{0.9804}\\
EMSRDPN-234 &$\times2$&\textbf{38.28}&\textbf{0.9660}&\textbf{34.33}&\textbf{0.9316}&\textbf{32.42}&\textbf{0.9122}&\textbf{33.34}&\textbf{0.9439}&\textbf{39.53}&\underline{0.9802}\\
EMSRDPN-2348 &$\times2$&38.22&\underline{0.9659}&\underline{34.26}&0.9309&\underline{32.40}&\underline{0.9120}&\underline{33.23}&\underline{0.9436}&39.46&0.9800\\
\hline
\hline
EMSRDPN-3 &$\times3$&34.67&0.9374&30.60&0.8614&29.26&0.8270&28.85&0.8762&34.29&0.9526\\
EMSRDPN-23 &$\times3$&34.69&0.9376&30.61&0.8624&29.30&0.8276&29.01&0.8790&34.52&0.9536\\
EMSRDPN-234 &$\times3$&\underline{34.77}&\textbf{0.9381}&30.68&\underline{0.8634}&\textbf{29.33}&\textbf{0.8284}&\underline{29.16}&\underline{0.8816}&\textbf{34.62}&\textbf{0.9541}\\
EMSRDPN-2348 &$\times3$&\textbf{34.80}&\underline{0.9380}&\textbf{30.71}&\textbf{0.8637}&29.31&\underline{0.8282}&\textbf{29.17}&\textbf{0.8818}&\underline{34.61}&\textbf{0.9541}\\
EMSRDPN-348 &$\times3$&\textbf{34.80}&\textbf{0.9381}&\underline{30.70}&0.8630&\underline{29.32}&0.8278&29.12&0.8810&34.53&\underline{0.9537}\\
\hline
\hline
EMSRDPN-4 &$\times4$&32.50&0.9083&28.88&0.8048&27.74&0.7607&26.78&0.8188&31.28&0.9221\\
EMSRDPN-48 &$\times4$&32.56&\underline{0.9090}&28.91&0.8059&27.78&0.7618&26.88&0.8220&31.36&0.9221\\
EMSRDPN-348 &$\times4$&\textbf{32.62}&\underline{0.9090}&\textbf{28.96}&\underline{0.8067}&\textbf{27.80}&\underline{0.7623}&\underline{26.98}&\underline{0.8246}&31.45&0.9232\\
EMSRDPN-2348 &$\times4$&\textbf{32.62}&\textbf{0.9093}&28.93&\textbf{0.8069}&\underline{27.79}&\textbf{0.7627}&\textbf{26.99}&\textbf{0.8253}&\underline{31.50}&\underline{0.9233}\\
EMSRDPN-234 &$\times4$&\underline{32.59}&\underline{0.9090}&\underline{28.95}&0.8063&\underline{27.79}&0.7622&26.96&0.8240&\textbf{31.51}&\textbf{0.9237}\\
\hline
\hline
EMSRDPN-8 &$\times8$&27.21&0.7827&25.15&0.6512&24.91&0.6040&22.77&0.6407&25.03&0.7958\\
EMSRDPN-48 &$\times8$&27.28&\underline{0.7844}&\underline{25.28}&\underline{0.6542}&\underline{24.95}&0.6055&\underline{23.00}&0.6494&25.27&0.8011\\
EMSRDPN-348 &$\times8$&\underline{27.32}&0.7839&25.25&0.6538&\textbf{24.96}&\underline{0.6058}&\textbf{23.05}&\underline{0.6517}&\textbf{25.32}&\underline{0.8020}\\
EMSRDPN-2348 &$\times8$&\textbf{27.34}&\textbf{0.7858}&\textbf{25.29}&\textbf{0.6547}&\textbf{24.96}&\textbf{0.6066}&\textbf{23.05}&\textbf{0.6525}&\underline{25.30}&\textbf{0.8025}\\
\hline
\end{tabular}
}
\end{center}
\label{tab:effect_of_multi_scale_pnsr_ssim_table}
}
\end{table*}

\begin{table*}[tb]
{
\fontsize{5}{6}\selectfont
\captionsetup{justification=centering}
\caption{Parameters and Inference Overhead for a $256\times256$ Image of EMSRDPN Using Different Number of Training Scale Factors. SSI: Single Scale Inference, MSI: Multiple Scale Inference.
}
\begin{center}
{
\begin{tabular}{|c|c|c|c|c|c|c|c|c|c|c|c|}
\hline
\multirow{3}*{Method}&\multirow{3}*{Params}&\multicolumn{5}{|c|}{Total Memory (MB)}&\multicolumn{5}{|c|}{Total Flops (T)}\\
\cline{3-12}
&&\multicolumn{4}{|c|}{SSI}&MSI&\multicolumn{4}{|c|}{SSI}&MSI\\
\cline{3-12}
&&$\times2$&$\times3$&$\times4$&$\times8$&$\times2$$\times3$$\times4$$\times8$&$\times2$&$\times3$&$\times4$&$\times8$&$\times2$$\times3$$\times4$$\times8$\\
\hline
\hline
EMSRDPN-2 &13,231,875&10627&\XSolidBrush&\XSolidBrush&\XSolidBrush&\XSolidBrush&0.87&\XSolidBrush&\XSolidBrush&\XSolidBrush&\XSolidBrush\\
EMSRDPN-3 &13,969,795&\XSolidBrush&10951&\XSolidBrush&\XSolidBrush&\XSolidBrush&\XSolidBrush&0.92&\XSolidBrush&\XSolidBrush&\XSolidBrush\\
EMSRDPN-4 &13,822,211&\XSolidBrush&\XSolidBrush&11660&\XSolidBrush&\XSolidBrush&\XSolidBrush&\XSolidBrush&1.03&\XSolidBrush&\XSolidBrush\\
EMSRDPN-8 &14,412,547&\XSolidBrush&\XSolidBrush&\XSolidBrush&15792&\XSolidBrush&\XSolidBrush&\XSolidBrush&\XSolidBrush&1.66&\XSolidBrush\\
\hline
\hline
EMSRDPN-2348 &17,522,188&10627&10951&11660&15792&17926&0.87&0.92&1.03&1.66&1.98\\
\hline
\end{tabular}
}
\end{center}
\label{tab:effect_of_multi_scale_overhead}
}
\end{table*}

\begin{figure}[t]
\setcounter{subfigure}{0}
  \centering
  \begin{subfigure}{.48\linewidth}
    \centering\includegraphics[width=\linewidth]{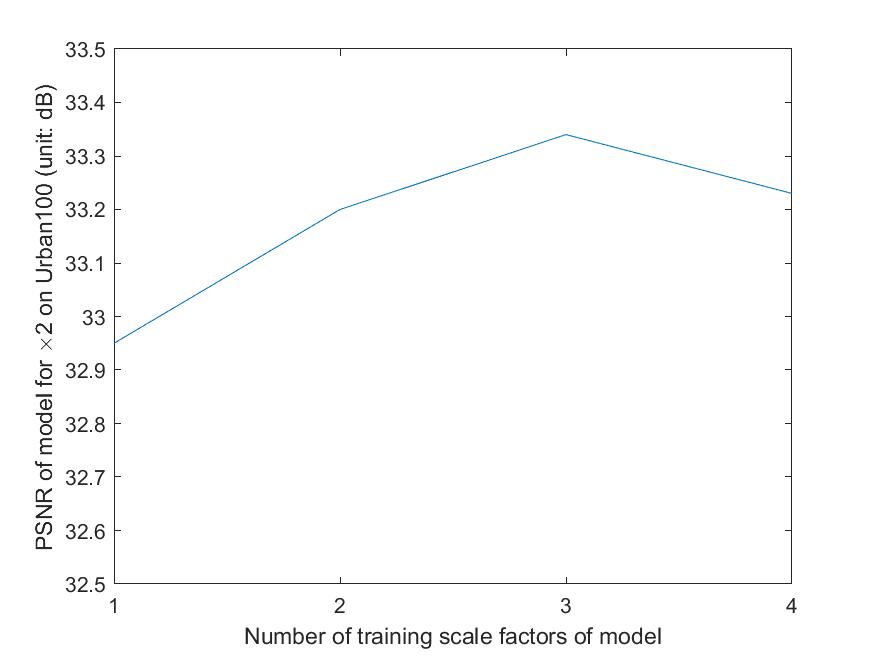}
    \caption{Performance trend of models for $\times2$}
  \end{subfigure}
  \begin{subfigure}{.48\linewidth}
    \centering\includegraphics[width=\linewidth]{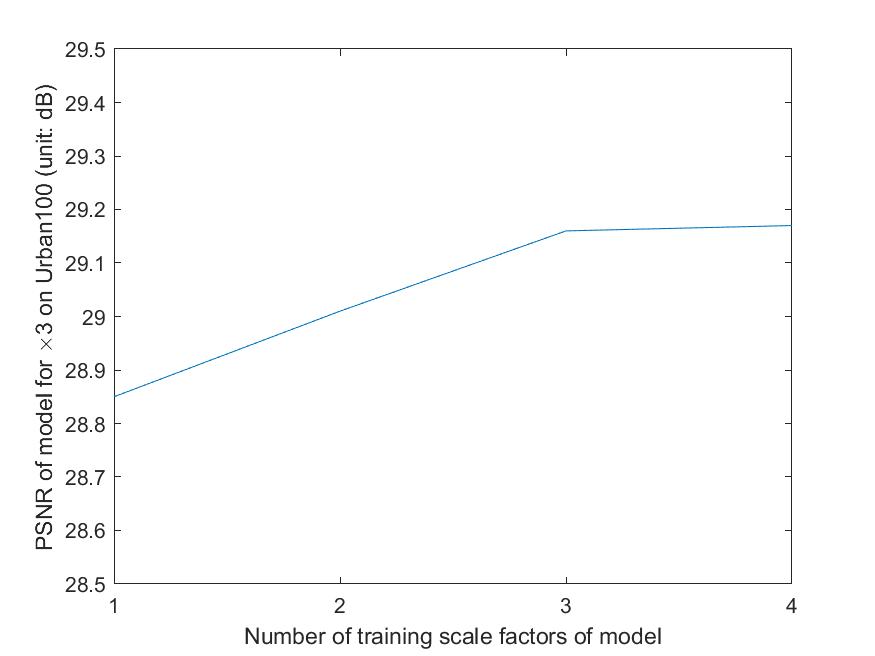}
    \caption{Performance trend of models for $\times3$}
  \end{subfigure}

  \begin{subfigure}{.48\linewidth}
    \centering\includegraphics[width=\linewidth]{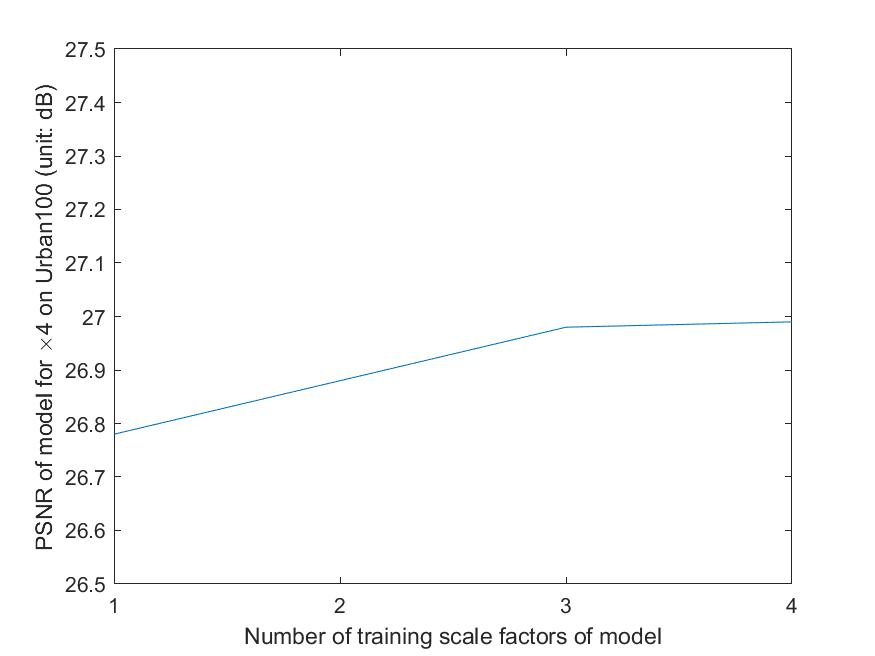}
    \caption{Performance trend of models for $\times4$}
  \end{subfigure}
  \begin{subfigure}{.48\linewidth}
    \centering\includegraphics[width=\linewidth]{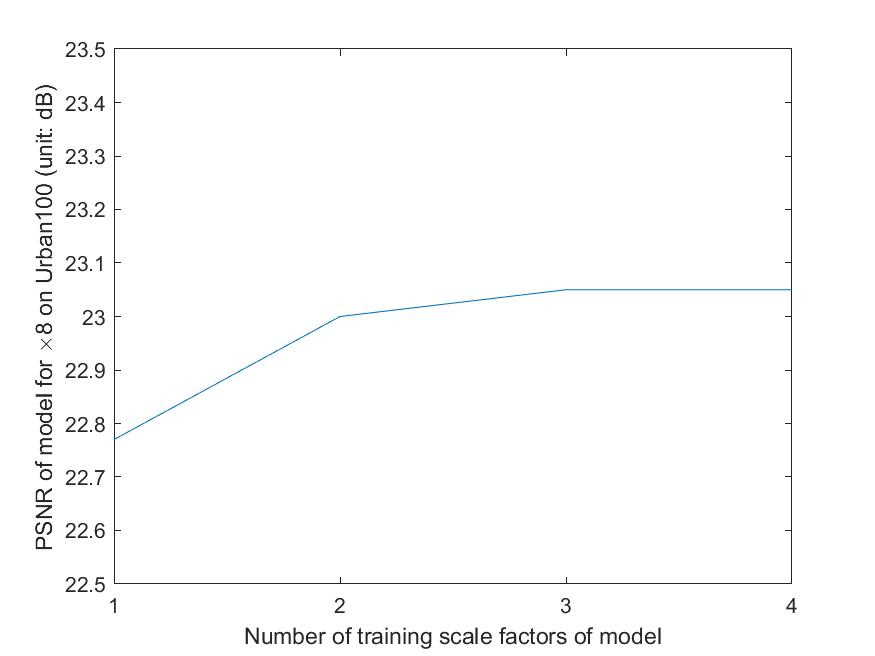}
    \caption{Performance trend of models for $\times8$}
  \end{subfigure}
  \caption{Illustration of effects to reconstruction performance of number of training scale factors.
  (a) Performance trend of models in group 2 for $\times2$.
  (b) Performance trend of models in group 2 for $\times3$.
  (c) Performance trend of models in group 3 for $\times4$.
  (d) Performance trend of models in group 3 for $\times8$.}
  \label{fig:DIV2K_Flickr2K_performance_trend_of_multi_scale_training}
\end{figure}
\subsection{Ablation Study}
\subsubsection{Effects of Multiple Scale Learning}
\label{sec:effects_of_multiple_scale_learning}
In this subsection, we study the effects to performance of multiple scale training and the benefits to efficiency of multiple scale inference of our EMSRDPN model. We design three groups of experiments to study the effects to performance indices of number of training scale factors. First, we train four models each with only one RB tail for a different specific scale factor in set of scale factors $\times2$, $\times3$, $\times4$ and $\times8$, denoted as EMSRDPN-2, EMSRDPN-3, EMSRDPN-4, EMSRDPN-8. Second, we start with single scale model of scale factor $\times2$, add one more scale factor each time according to the order of $\times3$, $\times4$, $\times8$ to train model of two scale factors, three scale factors, four scale factors, denoted as EMSRDPN-2, EMSRDPN-23, EMSRDPN-234, EMSRDPN-2348. Third, we start with single scale model of scale factor $\times8$, add one more scale factor each time according to the order of $\times4$, $\times3$, $\times2$ to train model of two scale factors, three scale factors, four scale factors, denoted as EMSRDPN-8, EMSRDPN-48, EMSRDPN-348, EMSRDPN-2348. The same models in different groups are trained only once.

The performance comparison are shown in Table \ref{tab:effect_of_multi_scale_pnsr_ssim_table} and Figure \ref{fig:DIV2K_Flickr2K_performance_trend_of_multi_scale_training}. We have two observations. First, $\times3$, $\times4$, $\times8$ all benefit from multiple scale training, the performance is increasing with the number of training scale factors. Second, $\times2$ benefits from multiple scale training up to three training scale factors, including EMSRDPN-23, EMSRDPN-234 models, but the performance indices drop a little for the model of four training scale factors, i.e. EMSRDPN-2348. We argue that it is because patterns of LR images of large scale factor of $\times8$ are much different from patterns of LR images of small scale factor of $\times2$ which deteriorates the feature correlation of these two scale factors.
According to these observations and analysis, we choose EMSRDPN-2348 as our final model abbreviated as EMSRDPN to compare to state-of-the-art SISR methods to achieve a good tradeoff between reconstruction performance and parameter and inference efficiency.

We also investigate the benefits to efficiency of multiple scale inference of our EMSRDPN model. As shown in Table \ref{tab:effect_of_multi_scale_overhead}, the multiple scale inference capability of our EMSRDPN model can amortize the parameters, memory consumption and computation flops between multiple scale factors. Our EMSRDPN model only add marginal parameters compared to a single scale model such as EMSRDPN-8, and only add marginal memory consumption and marginal computation flops to do inference of four scale factors compared to do inference of a single scale factor such as $\times8$ using EMSRDPN-8. The overhead of multiple scale inference of our EMSRDPN model is only fourth of overhead of multiple scale inference using single scale models.

\begin{table*}[tb]
{
\fontsize{5}{6}\selectfont
\captionsetup{justification=centering}
\caption{Average PSNR/SSIM Metrics for Scale Factor \(\times4\) of EMSRDPN-Res-4, EMSRDPN-Dense-4 and EMSRDPN-4. Best is \textbf{Highlighted}, and Second Best is \underline{Underlined.}
}
\begin{center}
{
\begin{tabular}{|c|c|c|c|c|c|c|c|c|c|c|c|}
\hline
\multirow{2}*{Method}&\multirow{2}*{Scale}&\multicolumn{2}{|c|}{Set5}&\multicolumn{2}{|c|}{Set14}&\multicolumn{2}{|c|}{B100}&\multicolumn{2}{|c|}{Urban100}&\multicolumn{2}{|c|}{Manga109}\\
\cline{3-12}
&&PSNR&SSIM&PSNR&SSIM&PSNR&SSIM&PSNR&SSIM&PSNR&SSIM\\
\hline
\hline
EMSRDPN-Res-4 &$\times4$&\textbf{32.52}&0.9082&\textbf{28.89}&\textbf{0.8050}&\underline{27.73}&0.7604&26.71&0.8179&\underline{31.17}&0.9213\\
EMSRDPN-Dense-4 &$\times4$&\textbf{32.52}&\textbf{0.9084}&28.87&\underline{0.8049}&\textbf{27.74}&\underline{0.7605}&\underline{26.76}&\underline{0.8186}&\textbf{31.28}&\underline{0.9215}\\
EMSRDPN-4 &$\times4$&\underline{32.50}&\underline{0.9083}&\underline{28.88}&0.8048&\textbf{27.74}&\textbf{0.7607}&\textbf{26.78}&\textbf{0.8188}&\textbf{31.28}&\textbf{0.9221}\\
\hline
\end{tabular}
}
\end{center}
\label{tab:effect_of_dual_path_connection_performance}
}
\end{table*}

\begin{table*}[tb]
{
\fontsize{5}{6}\selectfont
\captionsetup{justification=centering}
\caption{Network Configuration and Inference Overhead for a $256\times256$ Image of EMSRDPN-Res-4, EMSRDPN-Dense-4 and EMSRDPN-4.
SSI: Single Scale Inference, MSI: Multiple Scale Inference.
}
\begin{center}
{
\begin{tabular}{|c|c|c|c|c|c|c|c|c|c|c|c|}
\hline
\multirow{2}*{Method}&\multirow{2}*{$G_r$}&\multirow{2}*{$G_d$}&\multirow{2}*{$G$}&\multirow{2}*{$G_r+G_d$}&\multirow{2}*{$D$}&\multirow{2}*{$C$}&\multirow{2}*{Params}&\multicolumn{2}{|c|}{Total Memory (MB)}&\multicolumn{2}{|c|}{Total Flops (T)}\\
\cline{9-12}
&&&&&&&&SSI ($\times4$)&MSI&SSI ($\times4$)&MSI\\
\hline
\hline
EMSRDPN-Res-4 &135&0&0&135&16&4&13,915,803&10677&\XSolidBrush&1.05&\XSolidBrush\\
EMSRDPN-Dense-4 &0&122&122&122&16&4&13,748,671&12578&\XSolidBrush&1.01&\XSolidBrush\\
EMSRDPN-4 &64&64&64&128&16&4&13,822,211&11660&\XSolidBrush&1.03&\XSolidBrush\\
\hline
\end{tabular}
}
\end{center}
\label{tab:effect_of_dual_path_connection_overhead}
}
\end{table*}

\subsubsection{Effects of Dual Path Connections}
In this subsection, we study the effects of dual path connections proposed in our EMSRDPN model. Because residual connections and dense connections are special cases of the proposed dual path connections, we can simply set $G_d$ and $G$ to $0$ to form a model with a residual net structure or set $G_r$ to $0$ to form a model with a dense net structure. In this way, we can study the effects of dual path structure compared to pure residual net structure and dense net structure by limiting the models of three structures have similar number of parameters and similar number of network units thereof network depth. In this ablation study, we only train use models of different structure for $\times4$ scale factor. Table \ref{tab:effect_of_dual_path_connection_overhead} shows the network configuration of models of three structures, we use EMSRDPN-Res-4 to denote the model with only a residual net structure and EMSRDPN-Dense-4 to denote the model with only a dense net structure. Table \ref{tab:effect_of_dual_path_connection_performance} shows the performance comparisons for models of three structure. Table \ref{tab:effect_of_dual_path_connection_overhead} also includes the total memory and total flops of models with different structures during inference to compare memory and computation overheads.

As shown in Table \ref{tab:effect_of_dual_path_connection_performance}, EMSRDPN-4 has comparable even better performance compared to EMSRDPN-Res-4 and EMSRDPN-Dense-4. As shown in Table \ref{tab:effect_of_dual_path_connection_overhead}, the memory consumption of EMSRDPN-4 is less than EMSRDPN-Dense-4 and the computation flops of EMSRDPN-4 are less than EMSRDPN-Res-4.
These observations demonstrate dual path structure can achieve a good trade-off between performance, memory consumption and computation overhead.

\subsection{Comparisons with State-of-the-art Methods}
\begin{table*}[tb]
{
\fontsize{5}{6}\selectfont
\captionsetup{justification=centering}
\caption{Average PSNR/SSIM Metrics for Scale Factors \(\times2\), \(\times3\), \(\times4\), \(\times8\) of Compared Methods. Best is \textbf{Highlighted}, and Second Best is \underline{Underlined.}
}
\begin{center}
{
\begin{tabular}{|c|c|c|c|c|c|c|c|c|c|c|c|}
\hline
\multirow{2}*{Method}&\multirow{2}*{Scale}&\multicolumn{2}{|c|}{Set5}&\multicolumn{2}{|c|}{Set14}&\multicolumn{2}{|c|}{B100}&\multicolumn{2}{|c|}{Urban100}&\multicolumn{2}{|c|}{Manga109}\\
\cline{3-12}
&&PSNR&SSIM&PSNR&SSIM&PSNR&SSIM&PSNR&SSIM&PSNR&SSIM\\
\hline
\hline
Bicubic&$\times2$&33.66&0.9299&30.24&0.8688&29.56&0.8431&26.88&0.8403&30.80&0.9339\\
SRCNN \cite{dong2016image}&$\times2$&36.66&0.9542&32.45&0.9067&31.36&0.8879&29.50&0.8946&35.60&0.9663\\
FSRCNN \cite{dong2016accelerating}&$\times2$&37.05&0.9560&32.66&0.9090&31.53&0.8920&29.88&0.9020&36.67&0.9710\\
VDSR \cite{kim2016accurate}&$\times2$&37.53&0.9590&33.05&0.9130&31.90&0.8960&30.77&0.9140&37.22&0.9750\\
LapSRN \cite{lai2017deep}&$\times2$&37.52&0.9591&33.08&0.9130&31.08&0.8950&30.41&0.9101&37.27&0.9740\\
MemNet \cite{tai2017memnet}&$\times2$&37.78&0.9597&33.28&0.9142&32.08&0.8978&31.31&0.9195&37.72&0.9740\\
EDSR \cite{lim2017enhanced}&$\times2$&38.11&0.9602&33.92&0.9195&32.32&0.9013&32.93&0.9351&39.10&0.9773\\
SRMDNF \cite{DBLP:conf/cvpr/ZhangZ018}&$\times2$&37.79&0.9601&33.32&0.9159&32.05&0.8985&31.33&0.9204&38.07&0.9761\\
D-DBPN \cite{haris2018deep}&$\times2$&38.09&0.9600&33.85&0.9190&32.27&0.9000&32.55&0.9324&38.89&0.9775\\
RDN \cite{zhang2018residual}&$\times2$&38.24&0.9614&34.01&0.9212&32.34&0.9017&32.89&0.9353&39.18&0.9780\\
RCAN \cite{zhang2018image}&$\times2$&38.27&0.9614&34.12&0.9216&32.41&0.9027&33.34&0.9384&39.44&0.9786\\
NLRN \cite{DBLP:conf/nips/LiuWFLH18}&$\times2$&38.00&0.9603&33.46&0.9159&32.19&0.8992&31.81&0.9249&-&-\\
MSRN \cite{li2018multi}&$\times2$&38.08&0.9605&33.74&0.9170&32.23&0.9013&32.22&0.9326&38.82&\textbf{0.9868}\\
SAN \cite{DBLP:conf/cvpr/DaiCZXZ19}&$\times2$&\underline{38.31}&0.9620&34.07&0.9213&32.42&0.9028&33.10&0.9370&39.32&0.9792\\
HAN \cite{niu2020single}&$\times2$&38.27&0.9614&34.16&0.9217&32.41&0.9027&33.35&0.9385&39.46&0.9785\\
NLSN \cite{mei2021image}&$\times2$&\textbf{38.34}&0.9618&34.08&0.9231&\underline{32.43}&0.9027&\underline{33.42}&0.9394&\underline{39.59}&0.9789\\
EMSRDPN-D &$\times2$&38.23&0.9658&33.92&0.9289&32.33&0.9112&32.96&0.9408&39.11&0.9795\\
EMSRDPN &$\times2$&38.22&\underline{0.9659}&\underline{34.26}&\underline{0.9309}&32.40&\underline{0.9120}&33.23&\underline{0.9436}&39.46&0.9800\\
EMSRDPN+ &$\times2$&38.28&\textbf{0.9660}&\textbf{34.31}&\textbf{0.9317}&\textbf{32.44}&\textbf{0.9124}&\textbf{33.45}&\textbf{0.9448}&\textbf{39.61}&\underline{0.9804}\\
\hline
\hline
Bicubic&$\times3$&30.39&0.8682&27.55&0.7742&27.21&0.7385&24.46&0.7349&26.95&0.8556\\
SRCNN \cite{dong2016image}&$\times3$&32.75&0.9090&29.30&0.8215&28.41&0.7863&26.24&0.7989&30.48&0.9117\\
FSRCNN \cite{dong2016accelerating}&$\times3$&33.18&0.9140&29.37&0.8240&28.53&0.7910&26.43&0.8080&31.10&0.9210\\
VDSR \cite{kim2016accurate}&$\times3$&33.67&0.9210&29.78&0.8320&28.83&0.7990&27.14&0.8290&32.01&0.9340\\
LapSRN \cite{lai2017deep}&$\times3$&33.82&0.9227&29.87&0.8320&28.82&0.7980&27.07&0.8280&32.21&0.9350\\
MemNet \cite{tai2017memnet}&$\times3$&34.09&0.9248&30.00&0.8350&28.96&0.8001&27.56&0.8376&32.51&0.9369\\
EDSR \cite{lim2017enhanced}&$\times3$&34.65&0.9280&30.52&0.8462&29.25&0.8093&28.80&0.8653&34.17&0.9476\\
SRMDNF \cite{DBLP:conf/cvpr/ZhangZ018}&$\times3$&34.12&0.9254&30.04&0.8382&28.97&0.8025&27.57&0.8398&33.00&0.9403\\
RDN \cite{zhang2018residual}&$\times3$&34.71&0.9296&30.57&0.8468&29.26&0.8093&28.80&0.8653&34.13&0.9484\\
RCAN \cite{zhang2018image}&$\times3$&34.74&0.9299&30.65&0.8482&29.32&0.8111&29.09&0.8702&34.44&0.9499\\
NLRN \cite{DBLP:conf/nips/LiuWFLH18}&$\times3$&34.27&0.9266&30.16&0.8374&29.06&0.8026&27.93&0.8453&-&-\\
SAN \cite{DBLP:conf/cvpr/DaiCZXZ19}&$\times3$&34.75&0.9300&30.59&0.8476&29.33&0.8112&28.93&0.8671&34.30&0.9494\\
HAN \cite{niu2020single}&$\times3$&34.75&0.9299&30.67&0.8483&29.32&0.8110&29.10&0.8705&34.48&0.9500\\
NLSN \cite{mei2021image}&$\times3$&\textbf{34.85}&0.9306&30.70&0.8485&\underline{29.34}&0.8117&\underline{29.25}&0.8726&34.57&0.9508\\
EMSRDPN-D &$\times3$&34.74&0.9376&30.57&0.8619&29.26&0.8271&28.91&0.8775&34.24&0.9526\\
EMSRDPN &$\times3$&\underline{34.80}&\underline{0.9380}&\underline{30.71}&\underline{0.8637}&29.31&\underline{0.8282}&29.17&\underline{0.8818}&\underline{34.61}&\underline{0.9541}\\
EMSRDPN+ &$\times3$&\textbf{34.85}&\textbf{0.9384}&\textbf{30.78}&\textbf{0.8645}&\textbf{29.36}&\textbf{0.8289}&\textbf{29.31}&\textbf{0.8836}&\textbf{34.82}&\textbf{0.9550}\\
\hline
\hline
Bicubic&$\times4$&28.42&0.8104&26.00&0.7027&25.96&0.6675&23.14&0.6577&24.89&0.7866\\
SRCNN \cite{dong2016image}&$\times4$&30.48&0.8628&27.50&0.7513&26.90&0.7101&24.52&0.7221&27.58&0.8555\\
FSRCNN \cite{dong2016accelerating}&$\times4$&30.72&0.8660&27.61&0.7550&26.98&0.7150&24.62&0.7280&27.90&0.8610\\
VDSR \cite{kim2016accurate}&$\times4$&31.35&0.8830&28.02&0.7680&27.29&0.0726&25.18&0.7540&28.83&0.8870\\
LapSRN \cite{lai2017deep}&$\times4$&31.54&0.8850&28.19&0.7720&27.32&0.7270&25.21&0.7560&29.09&0.8900\\
MemNet \cite{tai2017memnet}&$\times4$&31.74&0.8893&28.26&0.7723&27.40&0.7281&25.50&0.7630&29.42&0.8942\\
EDSR \cite{lim2017enhanced}&$\times4$&32.46&0.8968&28.80&0.7876&27.71&0.7420&26.64&0.8033&31.02&0.9148\\
SRMDNF \cite{DBLP:conf/cvpr/ZhangZ018}&$\times4$&31.96&0.8925&28.35&0.7787&27.49&0.7337&25.68&0.7731&30.09&0.9024\\
D-DBPN \cite{haris2018deep}&$\times4$&32.47&0.8980&28.82&0.7860&27.72&0.7400&26.38&0.7946&30.91&0.9137\\
RDN \cite{zhang2018residual}&$\times4$&32.47&0.8990&28.81&0.7871&27.72&0.7419&26.61&0.8028&31.00&0.9151\\
RCAN \cite{zhang2018image}&$\times4$&32.63&0.9002&28.87&0.7889&27.77&0.7436&26.82&0.8087&31.22&0.9173\\
NLRN \cite{DBLP:conf/nips/LiuWFLH18}&$\times4$&31.92&0.8916&28.36&0.7745&27.48&0.7306&25.79&0.7729&-&-\\
MSRN \cite{li2018multi}&$\times4$&32.07&0.8903&28.60&0.7751&27.52&0.7273&26.04&0.7896&30.17&0.9034\\
SAN \cite{DBLP:conf/cvpr/DaiCZXZ19}&$\times4$&\underline{32.64}&0.9003&28.92&0.7888&27.78&0.7436&26.79&0.8068&31.18&0.9169\\
HAN \cite{niu2020single}&$\times4$&\underline{32.64}&0.9002&28.90&0.7890&\underline{27.80}&0.7442&26.85&0.8094&31.42&0.9177\\
NLSN \cite{mei2021image}&$\times4$&32.59&0.9000&28.87&0.7891&27.78&0.7444&26.96&0.8109&31.27&0.9184\\
EMSRDPN-D &$\times4$&32.50&0.9080&28.85&0.8052&27.74&0.7609&26.79&0.8198&31.20&0.9216\\
EMSRDPN &$\times4$&32.62&\underline{0.9093}&\underline{28.93}&\underline{0.8069}&27.79&\underline{0.7627}&\underline{26.99}&\underline{0.8253}&\underline{31.50}&\underline{0.9233}\\
EMSRDPN+ &$\times4$&\textbf{32.70}&\textbf{0.9101}&\textbf{29.02}&\textbf{0.8081}&\textbf{27.84}&\textbf{0.7637}&\textbf{27.14}&\textbf{0.8279}&\textbf{31.74}&\textbf{0.9252}\\
\hline
\hline
Bicubic&$\times8$&24.40&0.6580&23.10&0.5660&23.67&0.5480&20.74&0.5160&21.47&0.6500\\
SRCNN \cite{dong2016image}&$\times8$&25.33&0.6900&23.76&0.5910&24.13&0.5660&21.29&0.5440&22.46&0.6950\\
FSRCNN \cite{dong2016accelerating}&$\times8$&20.13&0.5520&19.75&0.4820&24.21&0.5680&21.32&0.5380&22.39&0.6730\\
SCN \cite{wang2015deep}&$\times8$&25.59&0.7071&24.02&0.6028&24.30&0.5698&21.52&0.5571&22.68&0.6963\\
VDSR \cite{kim2016accurate}&$\times8$&25.93&0.7240&24.26&0.6140&24.49&0.5830&21.70&0.5710&23.16&0.7250\\
LapSRN \cite{lai2017deep}&$\times8$&26.15&0.7380&24.35&0.6200&24.54&0.5860&21.81&0.5810&23.39&0.7350\\
MemNet \cite{tai2017memnet}&$\times8$&26.16&0.7414&24.38&0.6199&24.58&0.5842&21.89&0.5825&23.56&0.7387\\
MSLapSRN \cite{lai2018fast}&$\times8$&26.34&0.7530&24.57&0.6290&24.65&0.5920&22.06&0.5980&23.90&0.7590\\
EDSR \cite{lim2017enhanced}&$\times8$&26.96&0.7762&24.91&0.6420&24.81&0.5985&22.51&0.6221&24.69&0.7841\\
D-DBPN \cite{haris2018deep}&$\times8$&27.21&0.7840&25.13&0.6480&24.88&0.6010&22.73&0.6312&25.14&0.7987\\
RCAN \cite{zhang2018image}&$\times8$&27.31&\underline{0.7878}&25.23&0.6511&\underline{24.98}&0.6058&23.00&0.6452&25.24&\underline{0.8029}\\
MSRN \cite{li2018multi}&$\times8$&26.59&0.7254&24.88&0.5961&24.70&0.5410&22.37&0.5977&24.28&0.7517\\
SAN \cite{DBLP:conf/cvpr/DaiCZXZ19}&$\times8$&27.22&0.7829&25.14&0.6476&24.88&0.6011&22.70&0.6314&24.85&0.7906\\
HAN \cite{niu2020single}&$\times8$&27.33&\textbf{0.7884}&25.24&0.6510&\underline{24.98}&0.6059&22.98&0.6437&25.20&0.8011\\
EMSRDPN-D &$\times8$&27.28&0.7836&25.18&0.6514&24.93&0.6049&22.88&0.6456&25.09&0.7972\\
EMSRDPN &$\times8$&\underline{27.34}&0.7858&\underline{25.29}&\underline{0.6547}&24.96&\underline{0.6066}&\underline{23.05}&\underline{0.6525}&\underline{25.30}&0.8025\\
EMSRDPN+ &$\times8$&\textbf{27.41}&0.7873&\textbf{25.37}&\textbf{0.6564}&\textbf{25.01}&\textbf{0.6078}&\textbf{23.18}&\textbf{0.6566}&\textbf{25.50}&\textbf{0.8067}\\
\hline
\end{tabular}
}
\end{center}
\label{tab:DIV2K_Flickr2K_pnsr_ssim_table}
}
\end{table*}

\begin{figure*}[htb]
  \centering
  \begin{minipage}{.28\linewidth}
  \centering
  \begin{subfigure}{\linewidth}
    \centering\includegraphics[width=\linewidth]{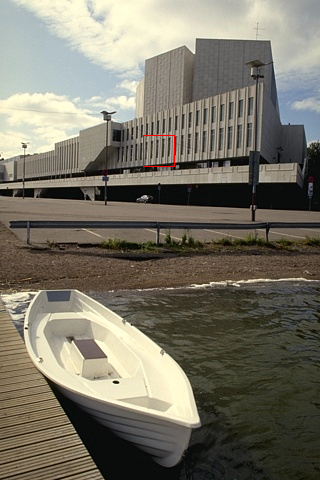}
    \caption*{B100 ($\times4$)\\78004}
  \end{subfigure}
  \end{minipage}
  \centering
  \begin{minipage}{.56\linewidth}
  \centering
  \begin{subfigure}{.24\linewidth}
    \centering\includegraphics[width=\linewidth]{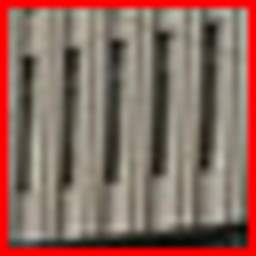}
    \caption*{HR\\PSNR/SSIM}
  \end{subfigure}
  \begin{subfigure}{.24\linewidth}
    \centering\includegraphics[width=\linewidth]{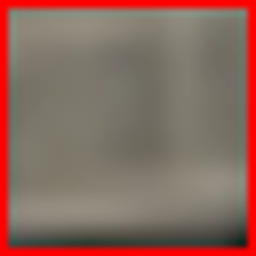}
	\caption*{Bicubic\\24.47/0.6480}
  \end{subfigure}
  \begin{subfigure}{.24\linewidth}
    \centering\includegraphics[width=\linewidth]{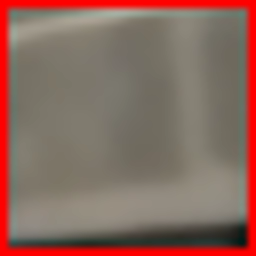}
	\caption*{LapSRN\\26.19/0.7422}
  \end{subfigure}
  \begin{subfigure}{.24\linewidth}
    \centering\includegraphics[width=\linewidth]{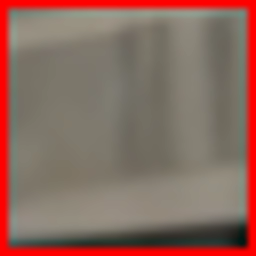}
	\caption*{MDSR\\27.00/0.7681}
  \end{subfigure}

  \begin{subfigure}{.24\linewidth}
    \centering\includegraphics[width=\linewidth]{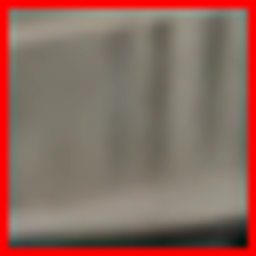}
	\caption*{D-DBPN\\26.92/0.7642}
  \end{subfigure}
  \begin{subfigure}{.24\linewidth}
    \centering\includegraphics[width=\linewidth]{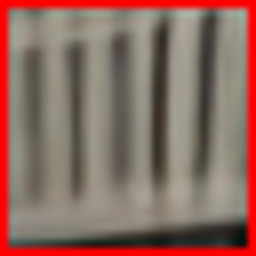}
	\caption*{RDN\\27.14/0.7755}
  \end{subfigure}
  \begin{subfigure}{.24\linewidth}
    \centering\includegraphics[width=\linewidth]{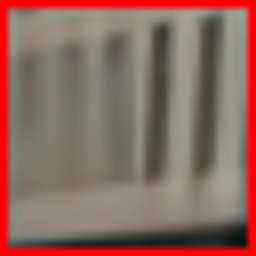}
	\caption*{RCAN\\27.40/0.7795}
  \end{subfigure}
  \begin{subfigure}{.24\linewidth}
    \centering\includegraphics[width=\linewidth]{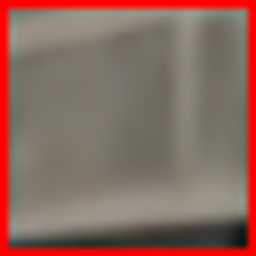}
	\caption*{MSRN\\26.92/0.7621}
  \end{subfigure}

  \begin{subfigure}{.24\linewidth}
    \centering\includegraphics[width=\linewidth]{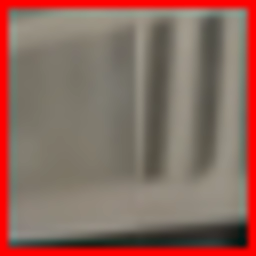}
	\caption*{SAN\\27.18/0.7748}
  \end{subfigure}
  \begin{subfigure}{.24\linewidth}
    \centering\includegraphics[width=\linewidth]{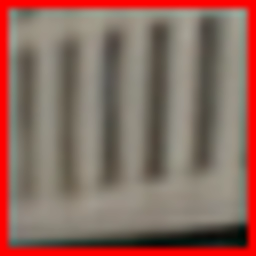}
	\caption*{HAN\\\underline{27.51}/\underline{0.7801}}
  \end{subfigure}
  \begin{subfigure}{.24\linewidth}
    \centering\includegraphics[width=\linewidth]{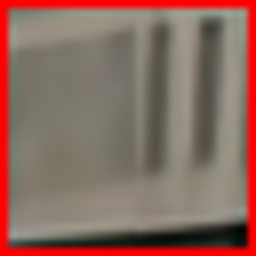}
	\caption*{NLSN\\27.29/0.7760}
  \end{subfigure}
  \begin{subfigure}{.24\linewidth}
    \centering\includegraphics[width=\linewidth]{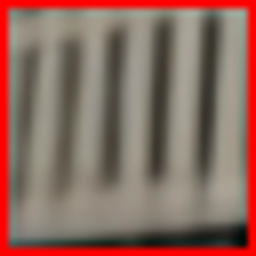}
	\caption*{EMSRDPN\\\textbf{27.51}/\textbf{0.7833}}
  \end{subfigure}
  \end{minipage}

  \centering
  \begin{minipage}{.28\linewidth}
  \centering
  \begin{subfigure}{\linewidth}
    \centering
    \includegraphics[width=\linewidth]{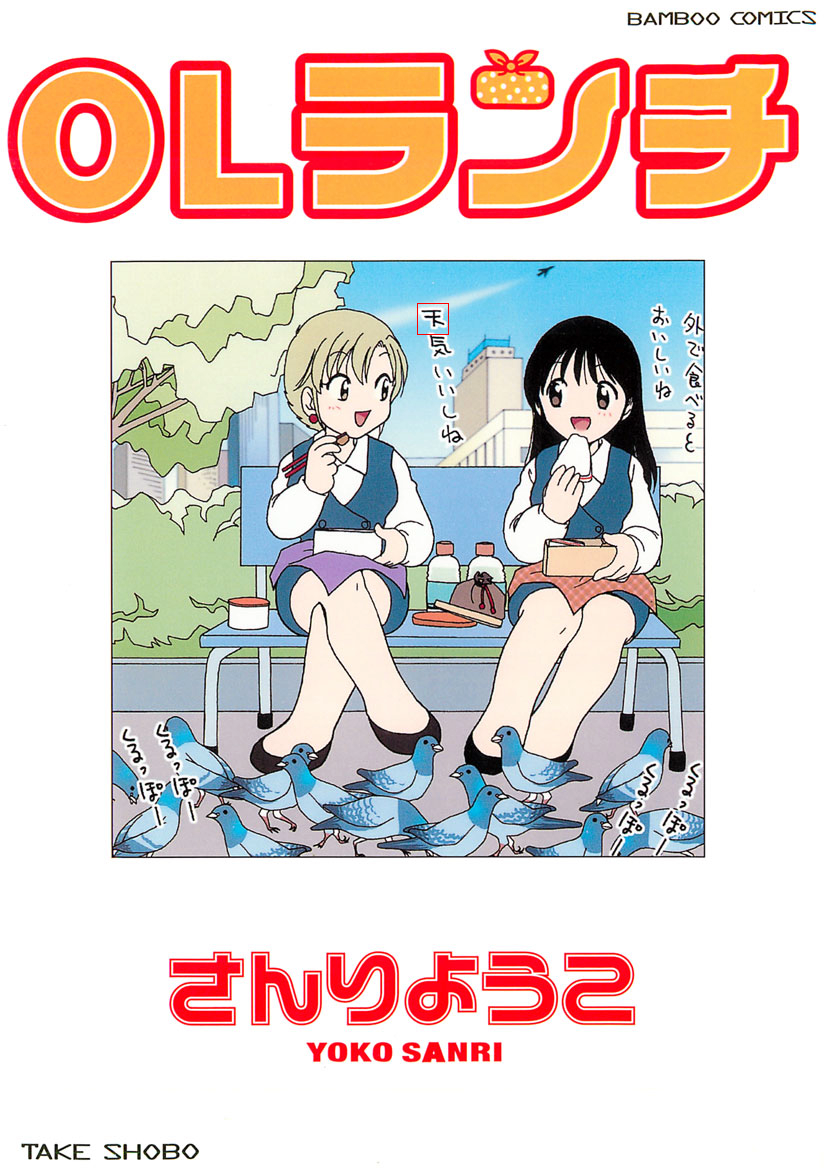}
    \caption*{Manga109 ($\times4$)\\OL\_Lunch}
  \end{subfigure}
  \end{minipage}
  \centering
  \begin{minipage}{.56\linewidth}
  \centering
  \begin{subfigure}{.24\linewidth}
    \centering
    \includegraphics[width=\linewidth]{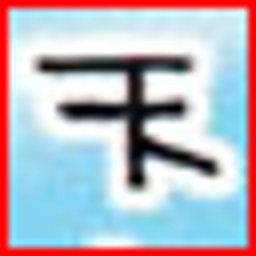}
    \caption*{HR\\PSNR/SSIM}
  \end{subfigure}
  \begin{subfigure}{.24\linewidth}
    \centering\includegraphics[width=\linewidth]{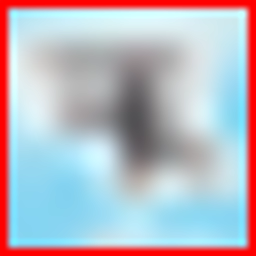}
	\caption*{Bicubic\\20.79/0.8335}
  \end{subfigure}
  \begin{subfigure}{.24\linewidth}
    \centering\includegraphics[width=\linewidth]{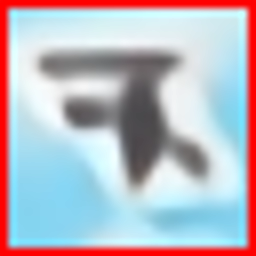}
	\caption*{LapSRN\\27.04/0.9449}
  \end{subfigure}
  \begin{subfigure}{.24\linewidth}
    \centering\includegraphics[width=\linewidth]{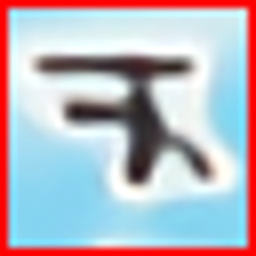}
	\caption*{MDSR\\28.09/0.9524}
  \end{subfigure}

  \begin{subfigure}{.24\linewidth}
    \centering\includegraphics[width=\linewidth]{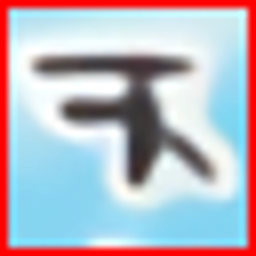}
	\caption*{D-DBPN\\27.45/0.9511}
  \end{subfigure}
  \begin{subfigure}{.24\linewidth}
    \centering\includegraphics[width=\linewidth]{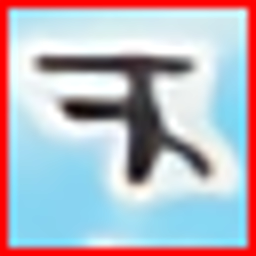}
	\caption*{RDN\\28.05/0.9527}
  \end{subfigure}
  \begin{subfigure}{.24\linewidth}
    \centering\includegraphics[width=\linewidth]{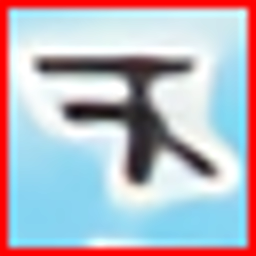}
	\caption*{RCAN\\27.96/0.9529}
  \end{subfigure}
  \begin{subfigure}{.24\linewidth}
    \centering\includegraphics[width=\linewidth]{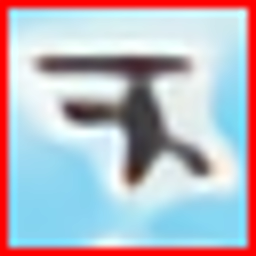}
	\caption*{MSRN\\27.51/0.9481}
  \end{subfigure}

  \begin{subfigure}{.24\linewidth}
    \centering\includegraphics[width=\linewidth]{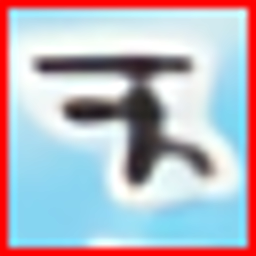}
	\caption*{SAN\\28.24/0.9537}
  \end{subfigure}
  \begin{subfigure}{.24\linewidth}
    \centering\includegraphics[width=\linewidth]{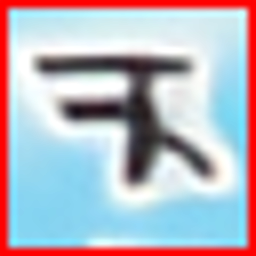}
	\caption*{HAN\\27.42/0.9473}
  \end{subfigure}
  \begin{subfigure}{.24\linewidth}
    \centering\includegraphics[width=\linewidth]{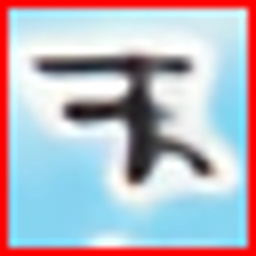}
	\caption*{NLSN\\\textbf{28.50}/\underline{0.9554}}
  \end{subfigure}
  \begin{subfigure}{.24\linewidth}
    \centering\includegraphics[width=\linewidth]{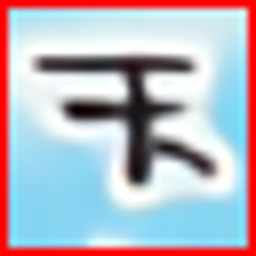}
	\caption*{EMSRDPN\\\underline{28.43}/\textbf{0.9560}}
  \end{subfigure}
  \end{minipage}
  \caption{Visual comparison for $\times4$ on B100 and Manga109 datasets.}
  \label{fig:visual_comparison_x4}
\end{figure*}

\begin{figure*}[htb]
  \centering
  \begin{minipage}{.28\linewidth}
  \centering
  \begin{subfigure}{\linewidth}
    \centering\includegraphics[width=\linewidth]{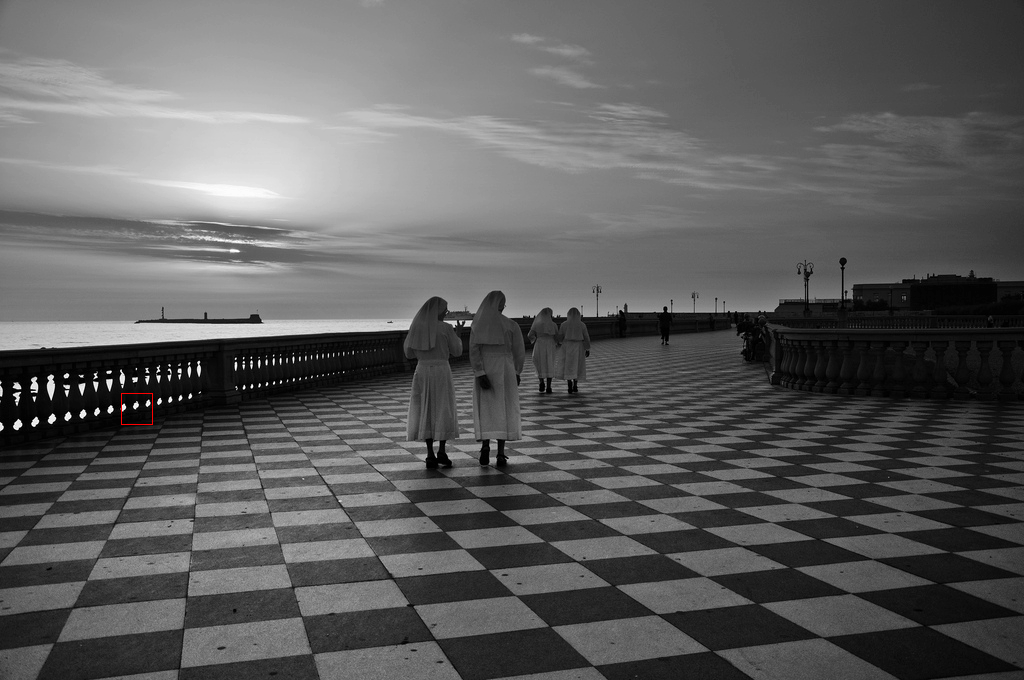}
    \caption*{Urban100 ($\times8$)\\img028}
  \end{subfigure}
  \end{minipage}
  \centering
  \begin{minipage}{.56\linewidth}
  \centering
  \begin{subfigure}{.24\linewidth}
    \centering\includegraphics[width=\linewidth]{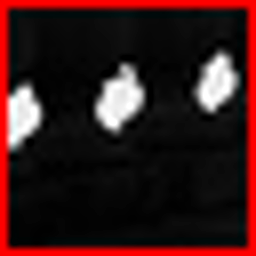}
    \caption*{HR\\PSNR/SSIM}
  \end{subfigure}
  \begin{subfigure}{.24\linewidth}
    \centering\includegraphics[width=\linewidth]{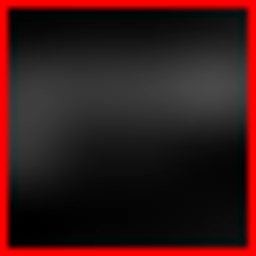}
	\caption*{Bicubic\\25.44/0.7179}
  \end{subfigure}
  \begin{subfigure}{.24\linewidth}
    \centering\includegraphics[width=\linewidth]{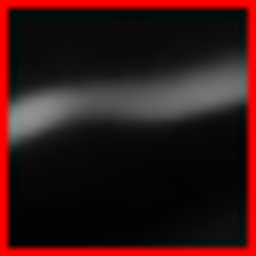}
	\caption*{LapSRN\\27.02/0.7787}
  \end{subfigure}
  \begin{subfigure}{.24\linewidth}
    \centering\includegraphics[width=\linewidth]{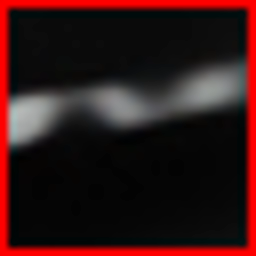}
	\caption*{D-DBPN\\27.14/0.7804}
  \end{subfigure}

  \begin{subfigure}{.24\linewidth}
    \centering\includegraphics[width=\linewidth]{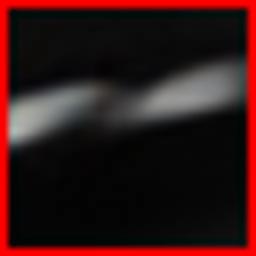}
	\caption*{RCAN\\\underline{27.74}/\underline{0.7961}}
  \end{subfigure}
  \begin{subfigure}{.24\linewidth}
    \centering\includegraphics[width=\linewidth]{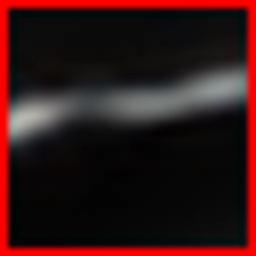}
	\caption*{MSRN\\27.16/0.7782}
  \end{subfigure}
  \begin{subfigure}{.24\linewidth}
    \centering\includegraphics[width=\linewidth]{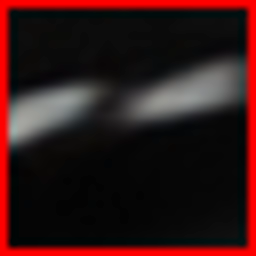}
	\caption*{HAN\\27.66/0.7948}
  \end{subfigure}
  \begin{subfigure}{.24\linewidth}
    \centering\includegraphics[width=\linewidth]{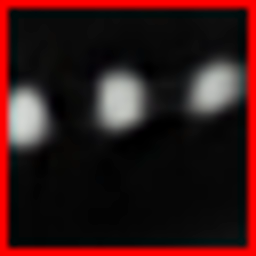}
	\caption*{EMSRDPN\\\textbf{28.07}/\textbf{0.8032}}
  \end{subfigure}
  \end{minipage}

  \centering
  \begin{minipage}{.28\linewidth}
  \centering
  \begin{subfigure}{\linewidth}
    \centering\includegraphics[width=\linewidth]{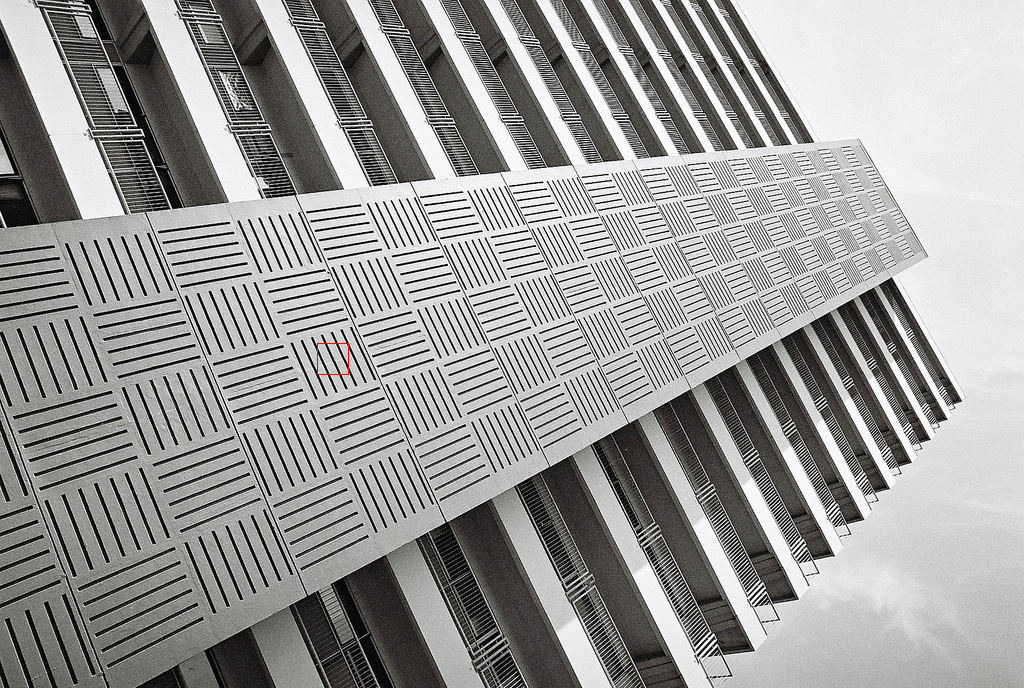}
    \caption*{Urban100 ($\times8$)\\img092}
  \end{subfigure}
  \end{minipage}
  \centering
  \begin{minipage}{.56\linewidth}
  \centering
  \begin{subfigure}{.24\linewidth}
    \centering\includegraphics[width=\linewidth]{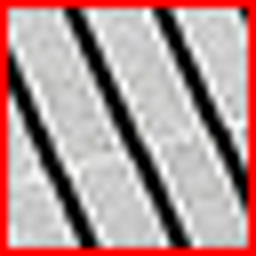}
    \caption*{HR\\PSNR/SSIM}
  \end{subfigure}
  \begin{subfigure}{.24\linewidth}
    \centering\includegraphics[width=\linewidth]{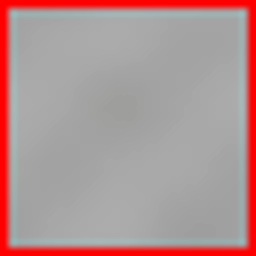}
	\caption*{Bicubic\\15.43/0.3268}
  \end{subfigure}
  \begin{subfigure}{.24\linewidth}
    \centering\includegraphics[width=\linewidth]{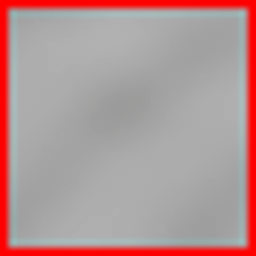}
	\caption*{LapSRN\\15.80/0.3983}
  \end{subfigure}
  \begin{subfigure}{.24\linewidth}
    \centering\includegraphics[width=\linewidth]{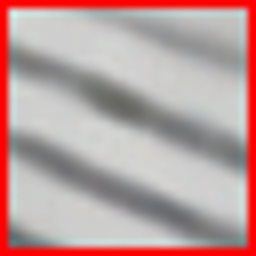}
	\caption*{D-DBPN\\16.18/0.4417}
  \end{subfigure}

  \begin{subfigure}{.24\linewidth}
    \centering\includegraphics[width=\linewidth]{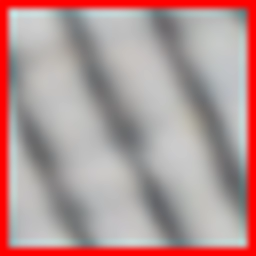}
	\caption*{RCAN\\16.35/0.4374}
  \end{subfigure}
  \begin{subfigure}{.24\linewidth}
    \centering\includegraphics[width=\linewidth]{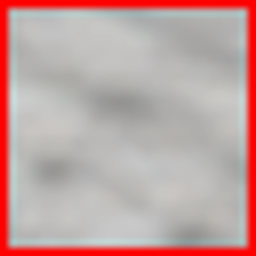}
	\caption*{MSRN\\15.83/0.4002}
  \end{subfigure}
  \begin{subfigure}{.24\linewidth}
    \centering\includegraphics[width=\linewidth]{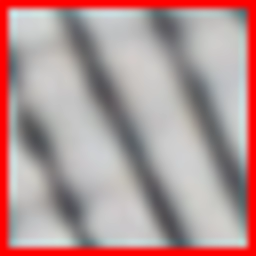}
	\caption*{HAN\\\underline{16.38}/\underline{0.4420}}
  \end{subfigure}
  \begin{subfigure}{.24\linewidth}
    \centering\includegraphics[width=\linewidth]{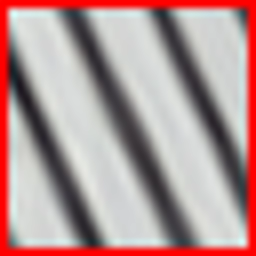}
	\caption*{EMSRDPN\\\textbf{16.74}/\textbf{0.4771}}
  \end{subfigure}
  \end{minipage}

  \centering
  \begin{minipage}{.28\linewidth}
  \centering
  \begin{subfigure}{\linewidth}
    \centering\includegraphics[width=\linewidth]{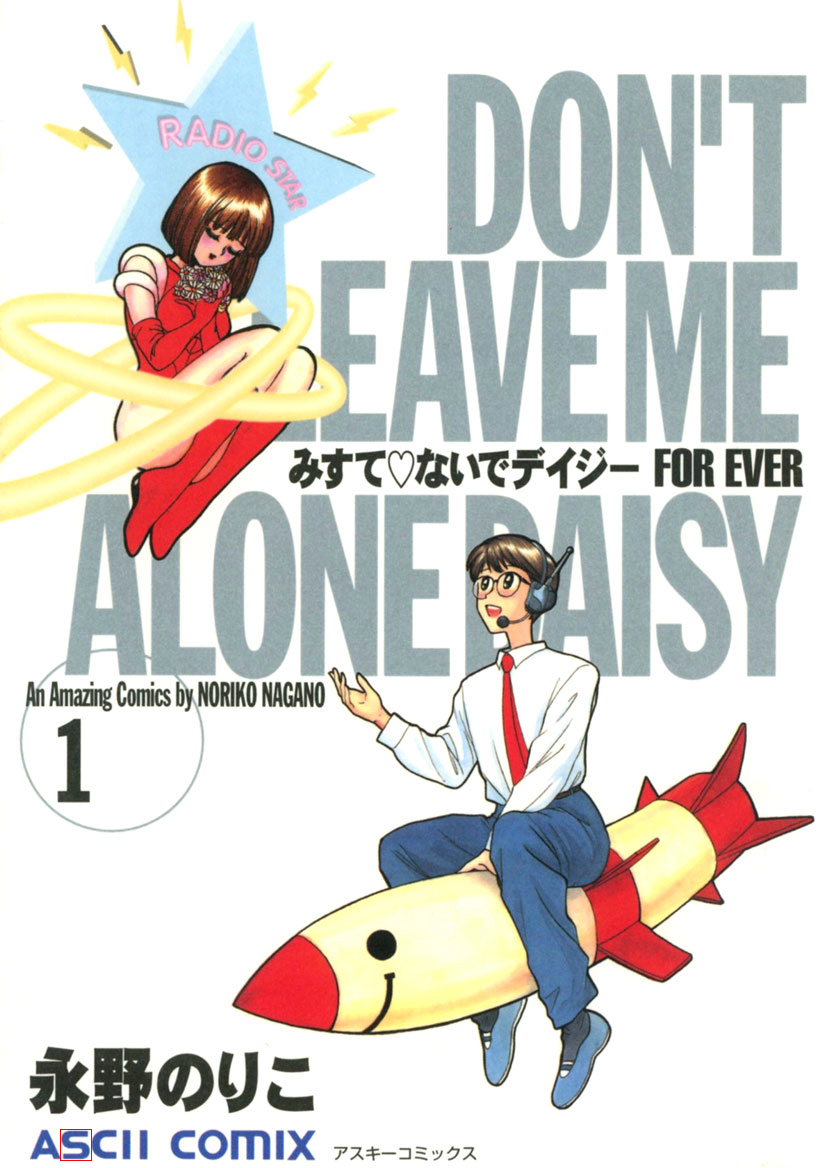}
    \caption*{Manga109 ($\times8$)\\MisutenaideDaisy}
  \end{subfigure}
  \end{minipage}
  \centering
  \begin{minipage}{.56\linewidth}
  \centering
  \begin{subfigure}{.24\linewidth}
    \centering\includegraphics[width=\linewidth]{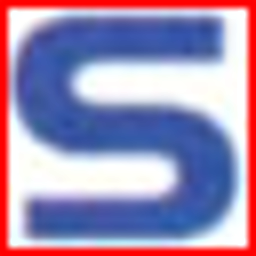}
    \caption*{HR\\PSNR/SSIM}
  \end{subfigure}
  \begin{subfigure}{.24\linewidth}
    \centering\includegraphics[width=\linewidth]{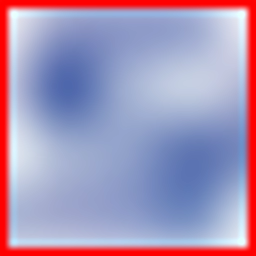}
	\caption*{Bicubic\\22.61/0.7747}
  \end{subfigure}
  \begin{subfigure}{.24\linewidth}
    \centering\includegraphics[width=\linewidth]{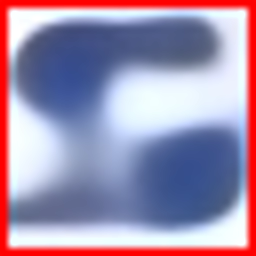}
	\caption*{LapSRN\\26.14/0.9018}
  \end{subfigure}
  \begin{subfigure}{.24\linewidth}
    \centering\includegraphics[width=\linewidth]{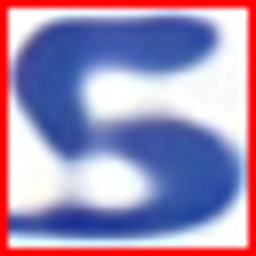}
	\caption*{D-DBPN\\27.33/0.9242}
  \end{subfigure}

  \begin{subfigure}{.24\linewidth}
    \centering\includegraphics[width=\linewidth]{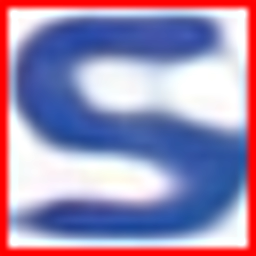}
	\caption*{RCAN\\\underline{27.58}/\underline{0.9254}}
  \end{subfigure}
  \begin{subfigure}{.24\linewidth}
    \centering\includegraphics[width=\linewidth]{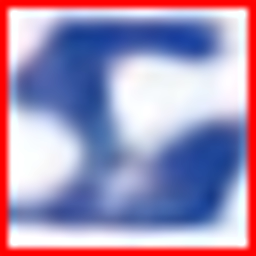}
	\caption*{MSRN\\26.62/0.9093}
  \end{subfigure}
  \begin{subfigure}{.24\linewidth}
    \centering\includegraphics[width=\linewidth]{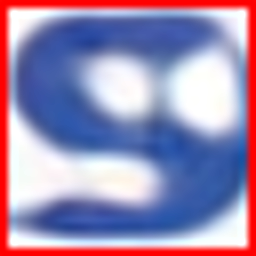}
	\caption*{HAN\\27.49/0.9245}
  \end{subfigure}
  \begin{subfigure}{.24\linewidth}
    \centering\includegraphics[width=\linewidth]{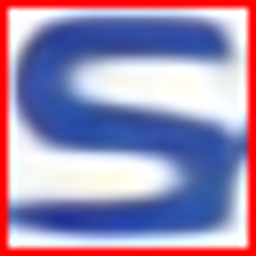}
	\caption*{EMSRDPN\\\textbf{27.65}/\textbf{0.9269}}
  \end{subfigure}
  \end{minipage}
  \caption{Visual comparison for $\times8$ on Urban100 and Manga109 datasets.}
  \label{fig:visual_comparison_x8}
\end{figure*}

In this subsection, we compare our method with SRCNN \cite{dong2016image}, FSRCNN \cite{dong2016accelerating}, VDSR \cite{kim2016accurate}, LapSRN \cite{lai2017deep}, MemNet \cite{tai2017memnet}, EDSR \cite{lim2017enhanced}, SRMDNF \cite{DBLP:conf/cvpr/ZhangZ018}, D-DBPN \cite{haris2018deep}, RDN \cite{zhang2018residual}, RCAN \cite{zhang2018image}, MSRN \cite{li2018multi}, NLRN \cite{DBLP:conf/nips/LiuWFLH18}, SAN \cite{DBLP:conf/cvpr/DaiCZXZ19}, HAN \cite{niu2020single} and NLSN \cite{mei2021image} to validate the effectiveness of EMSRDPN. Following the literature, self-ensemble strategy \cite{lim2017enhanced} is also adopted to improve EMSRDPN and self-ensembled EMSRDPN is denoted as EMSRDPN+. Table \ref{tab:DIV2K_Flickr2K_pnsr_ssim_table} shows the peak signal-to-noise ratio (PSNR) and the structural similarity index measure (SSIM) metrics on five benchmark datasets. Our method achieves comparable even better performance over state-of-the-art methods on almost all combinations of benchmark datasets and scale factors.

Figure \ref{fig:visual_comparison_x4}, \ref{fig:visual_comparison_x8} show visual comparisons. As shown in image \(78004\) from B100, images \(img028\) and \(img092\) from Urban100, our method reconstructs more textures, less blurring and ring effects of images than other state-of-the-art methods. As shown in images \(OL Lunch\) and \(MisutenaideDaisy\) from Manga109, our method reconstructs clearer edges and less artifacts of images than other state-of-the-art methods. Visual comparisons show our method reconstructs more high frequency details such as edges and textures of images and has less artifacts.

\subsection{Model Complexity and Inference Time}
\begin{figure}[tb]
\setcounter{subfigure}{0}
  \centering
  \begin{subfigure}{.48\linewidth}
    \centering\includegraphics[width=\linewidth]{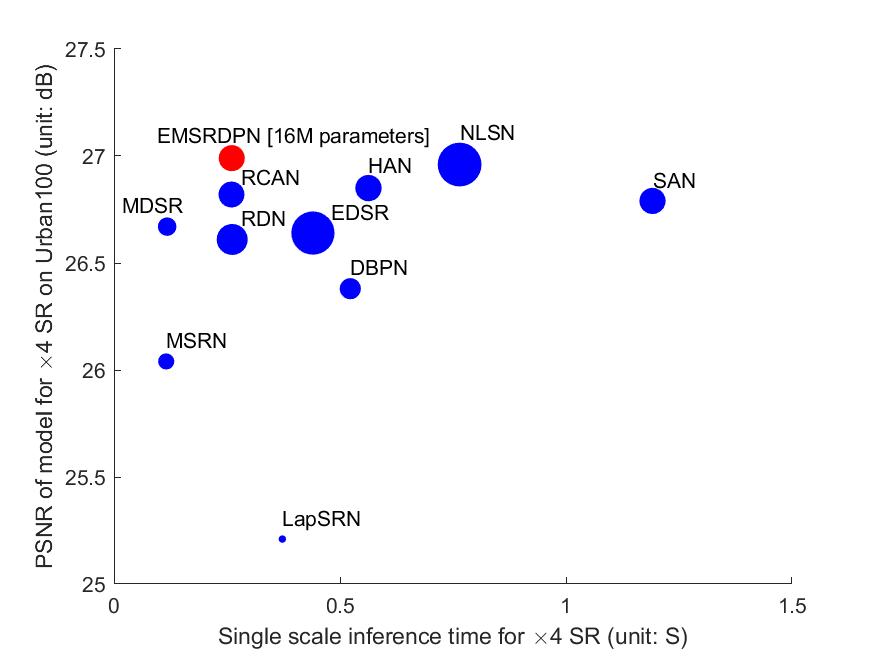}
    \caption{Comparison for single scale inference}
  \end{subfigure}
  \begin{subfigure}{.48\linewidth}
    \centering\includegraphics[width=\linewidth]{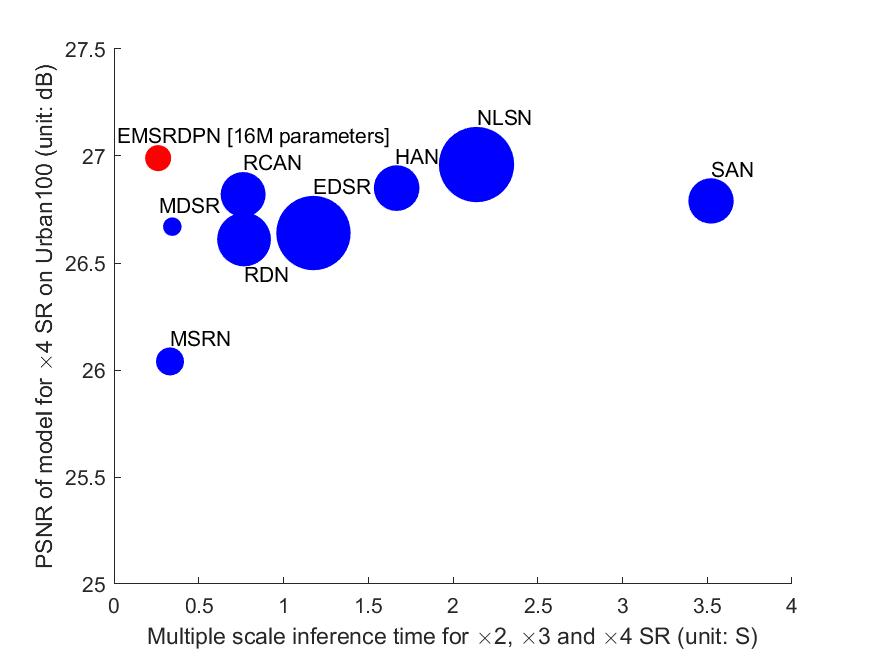}
    \caption{Comparison for multiple scale inference}
  \end{subfigure}
  \caption{Comparison of the performance, number of parameters and inference time of models, the inference time is tested using a $256\times256$ image on a NVIDIA TITAN Xp GPU. (a) Comparison for single scale inference of $\times4$, the size of solid circle is proportional to number of parameters of model for $\times4$. (b) Comparison for multiple scale inference of $\times2$, $\times3$, and $\times4$, the size of solid circle is proportional to number of parameters of models for $\times2$, $\times3$ and $\times4$.}
  \label{fig:DIV2K_performance_inference_time_parameters}
\end{figure}

In this subsection, we compare number of parameters and inference time of our EMSRDPN model with state-of-the-art methods. For single scale inference such as $\times4$ as shown in Figure \ref{fig:DIV2K_performance_inference_time_parameters}(a), our method has more parameters or inference time than LapSRN \cite{lai2017deep}, MDSR \cite{lim2017enhanced}, MSRN \cite{li2018multi} and DBPN \cite{haris2018deep} but it has much better reconstruction performance, meanwhile, our method has comparable even much less parameters and inference time than other state-of-the-art methods except that it has better reconstruction performance than these state-of-the-art methods. The efficiency of our method is much obvious especially for multiple scale inference such as $\times2$, $\times3$, and $\times4$ as shown in Figure \ref{fig:DIV2K_performance_inference_time_parameters}(b), our method has comparable parameters and inference time to MDSR \cite{lim2017enhanced} and MSRN \cite{li2018multi} and much less parameters and inference time than other state-of-the-art methods, but it has much better reconstruction performance than MDSR \cite{lim2017enhanced} and MSRN \cite{li2018multi} and better reconstruction performance than other state-of-the-art methods.
It is worth noting that the additional multiple scale inference time of $\times2$, $\times3$, and $\times4$ is only marginal compared to the single scale inference time of $\times4$, because the most computation can be shared between different scale factors during inference, which is consistent with results in Table \ref{tab:effect_of_multi_scale_overhead} and Section \ref{sec:effects_of_multiple_scale_learning}.
The statistics of number of parameters and inference time in this subsection and reconstruction performance in previous subsection show that our method can achieve a good trade-off between number of parameters, inference time and reconstruction performance.

\section{Conclusion and Future Work}
In this work,
we propose an efficient single image super-resolution network using dual path connections with multiple scale learning named as EMSRDPN to boost reconstruction performance, save parameters and improve computation efficiency for SISR. First, we introduce efficient dual path connections into a very deep convolutional neural network, which have the benefits of both residual connections to reuse common features and dense connections to explore new features to learn a good representation for SISR. Second, based on the improved network architecture to do SISR in LR space to save computation and memory cost, we propose a multiple scale training and inference architecture of network by sharing most of network parameters between different scale factors to make use of training data of multiple scale factors for each other to utilize feature correlation during training and amortize most parameters and computation between multiple scale factors during inference, which has benefits of both achieving better performance and improving parameter efficiency and inference time of network. Experiments show that our new model achieves better performance and improved visual effects can be seen in the results compared to state-of-the-art methods. Meanwhile, our method has comparable or even better parameter efficiency and inference time compared to state-of-the-art methods. We believe the proposed multiple scale training and inference strategy is generally applicable to various network designs to boost performance and efficiency of SISR and dual path connections could also be useful to other multimedia tasks than SISR.
\bibliographystyle{ACM-Reference-Format}
\bibliography{egbib}

\appendix

\end{document}